\documentclass[final,lp]{elsarticle}
\usepackage{lineno,hyperref}
\modulolinenumbers[5]
\usepackage{graphics}
\usepackage{dcolumn}
\usepackage{bm}   
\usepackage{color}
\usepackage{amssymb}
\usepackage{amsmath}
\usepackage{dsfont}
\usepackage{ulem}
\usepackage{physics}
\usepackage{dsfont}
\usepackage{subcaption}
\usepackage[justification=raggedright,labelfont={bf}]{caption}

\journal{Physics Letters A}

\begin{document}

\begin{frontmatter}

\title{Dissipative optical solitons in asymmetric Rosen-Morse potential}

\author[nit]{K.~Hari}
\author[nit]{\corref{cor1}K.~Manikandan}
\ead{manikandan.cnld@gmail.com}
\author[nit]{R. Sankaranarayanan}
\ead{sankar@nitt.edu}
\address[nit]{Department of Physics, National Institute of Technology, Tiruchirappalli - 620015, Tamil Nadu, India.}
\cortext[cor1]{Corresponding author;}

\date{\today}
\begin{abstract}
We investigate the existence and stability of dissipative soliton solution in a system described by complex Ginzburg-Landau (CGL) equation with asymmetric complex potential, which is obtained from original parity reflection - time reversal ($\mathcal{PT}$) symmetric Rosen-Morse potential. In this study, stability of solution is examined by numerical analysis to show that solitons are stable for some parameter ranges for both self-focusing and self-defocusing nonlinear modes. Dynamical properties such as evolution and transverse energy flow for both modes are also analyzed.  Obtained results are useful for experimental designs and applications in related fields. 
\end{abstract}

\end{frontmatter}

\section{Introduction}
Nonlinear systems have always been attracting researchers for many decades due to their rich dynamical features.  In recent years, many experimental implications have accelerated investigations on such systems to many fold.  Solitons are a class of experimentally realizable solutions for nonlinear differential equations. They are localized structures and have particle like properties. In particular, solitons can be realized as profile of light pulse in optical system, Bose-Einstein condensates (BECs), superconductivity, metamaterials, etc. Nonlinear systems can be either closed (conservative) or open (dissipative).  In open systems, we have to account for loss/gain of energy while investigating their solutions. In such a system, soliton solution can be stable even if the system is not in equilibrium due to the balance between loss/gain of energy.  

Many studies on optical systems can be carried out using complex Ginzburg-Landau (CGL) equation which accommodates dissipative solitons.  Existence of dissipative solitons is also guaranteed by the balance between nonlinearity and dispersion \cite{1, 2}. Such a balance in the system is characterized by system parameters and not by initial conditions. This is an interesting property since most of the nonlinear systems hugely depend on the initial conditions where even a small deviation leads to instability. This property is significant in experimental realizations where system parameters are used to identify the region for stable solitons. Since solitons do exist only in certain values of system parameters, dissipative system does not permit continuous families of solitons parameterized by propagation constant \cite{3}. 

Existence of dissipative nonlinear modes in complex-valued optical potentials described by CGL equation is reported elsewhere \cite{4,5}. It is worth noting that dissipative solitons have been implemented in semiconductor optical amplifiers \cite{6}, semiconductor resonators \cite{7}, semiconductor micro-cavities \cite{8}. In Ref. \cite{1}, many theoretical works on dissipative systems are consolidated. Another class of nonlinear system which is  described by nonlinear Schr\"{o}dinger (NLS) equation also supports dissipative solitons. A notable reference by F. K. Abdullaev et al. \cite{9} studies exact solutions for NLS equation with complex linear and nonlinear potentials.

In parity and time reversal $(\mathcal{PT})-$symmetric systems, originated from Bender and Boettcher \cite{10} the scenario is different, where continuous families of solitons exists. Many theoretical studies \cite{11,12} and experimental realizations \cite{13} have been reported in such systems.  $\mathcal{PT}-$symmetry breaking is also an important physical phenomenon because some of the desired properties are achieved at the point of symmetry breaking \cite{14} and in the symmetry broken phase \cite{15,16,17}. Theoretical studies on existence of stable soliton solutions in non$-\mathcal{PT}$ symmetric systems are also reported \cite{18,19,20,20a}. Existence of continuous families of soliton is also reported in such non$-\mathcal{PT}$ symmetric cases \cite{18,19}. 

It has been shown that $\mathcal{PT}-$symmetric system described by NLS equation with Rosen-Morse potential does not support any stable solutions \cite{12}.  Here we have tried to explore the possibility of stable solutions by modifying the Rosen-Morse potential and applying it in CGL system.  In light of the above, here, we consider a dissipative system where the governing equation is CGL equation with modified Rosen-Morse potential. The additional terms in CGL equation and the potential offer us the extra freedom to create new stable soliton solution in this system. Our analysis shows that the values of parameter regions for stable solutions are isolated rather than continuous.  This is in sharp contrast with stable solitons exist in  near $\mathcal{PT}-$symmetric Scarf-II potentials of CGL equation \cite{21}. 

This paper is organized as follows: In Sec. 2, we present a physical model of the system where we have also included the nature of original Rosen-Morse potential. In Sec. 3, we study the self-focusing nonlinear mode which contains nature of the potential, exact solution, stability analysis for the exact solution, soliton evolution and energy flow in the system. In Sec. 4, we extend our investigation for self-defocusing nonlinear mode. Finally in Sec. 5 we present a summary of results with conclusion.

\section{Physical model of the system}
Transmission of dissipative soliton in an optical lattice is given by CGL equation \cite{21,22,23} with complex constants and potentials as
\begin{align}\label{eq1}
i\frac{\partial\Psi}{\partial z}+(\alpha_{1}+i \alpha_{2})\frac{\partial^{2}\Psi}{\partial x^{2}}+[V(x)+i W(x)] \Psi + \sigma(\beta_{1}+i \beta_{2})|\Psi|^{2} \Psi = 0,
\end{align} 
where $\Psi(x,z)$ is the amplitude of complex electric field of optical pulse, $z$ is the propagation distance and $x$ is the spatial coordinate.  The parameters of the optical system are as given below:  $\alpha_{1}$ - diffraction coefficient, $\beta_{1}$ - Kerr-nonlinearity coefficient, $\alpha_{2}$ - spectral filtering, $\beta_{2}$ - nonlinear gain/loss in the system and $\sigma=\pm 1$ signifies self-focusing and self-defocusing nonlinear mode.  Here $V(x)$ and $W(x)$ are the real and imaginary parts of the potential, respectively. It is noted that the CGL equation is non$-\mathcal{PT}$ invariant due to the presence of complex coefficients. 

Our analysis starts by assuming stationary solution of the form $\Psi(x,z)=\phi(x)e^{i\mu z}$, where $\mu$ is the real valued propagation constant and $\phi(x)$ is the complex field. Substituting this solution into Eq. (\ref{eq1}), we obtain the following second-order ordinary differential equation of the form for $\mu=\alpha_{1}$, as
\begin{align}\label{eq2}
(\alpha_{1}+ i\alpha_{2}) \frac{d^{2}\phi}{dx^{2}} +[V(x)+iW(x)]\phi  +\sigma(\beta_{1}+i\beta_{2})|\phi|^{2}\phi-\mu \phi=0.
\end{align}

The general form of original $\mathcal{PT}-$symmetric Rosen-Morse potential \cite{16} is given by 
\begin{subequations}\label{eq3}
\begin{align}
\label{eq3a} V(x) = & A(A+1) \sech^{2}(x), \\
\label{eq3b} W(x) = & 2B \tanh (x)
\end{align}
\end{subequations}
where $A$ and $B$ describe the strength of real and imaginary parts of the potential, respectively. It is trivial to infer the shape of potential wherein $V(x)$ is an even function and $W(x)$ is an odd function. In what follows, we discuss a modified potentials for self focusing and self defocusing nonlinear modes. 
\section{Self-focusing nonlinear mode}
\subsection{Nature of potential}
In order to investigate the optical beam propagation of (\ref{eq1}) in self-focusing mode, we consider the complex asymmetric potential of the form
\begin{subequations}\label{eq4}
\begin{align}
\label{eq4a} V(x) = & -a(a+1)\sech^{2}(x)-2\left(\frac{\alpha_2}{\alpha_1}\right)b\tanh(x)+\frac{b^2}{\alpha_1}, \\
\label{eq4b} W(x) = & 2b\tanh(x)+W_1\sech^2(x)+\alpha_2\left(\frac{b^2}{\alpha_1^2}-1\right),
\end{align}
\end{subequations}
\begin{figure*}[!ht]
\centering
\begin{subfigure}[b]{0.325\textwidth}
\includegraphics[width=1\textwidth]{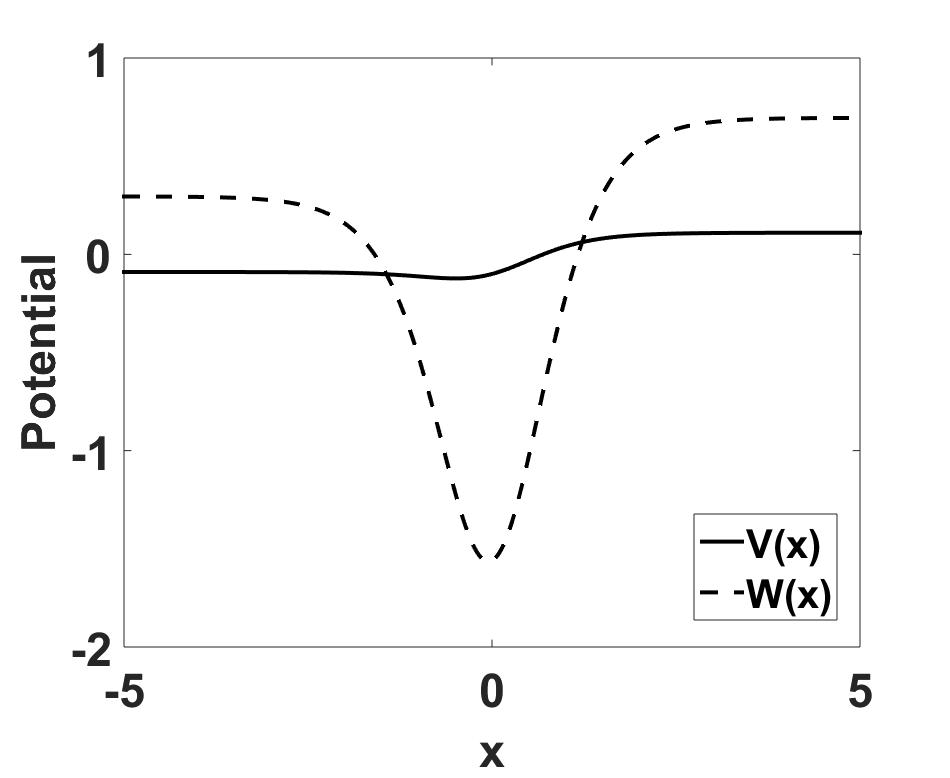}
\caption{}
\end{subfigure}
\begin{subfigure}[b]{0.325\textwidth}
\includegraphics[width=1\textwidth]{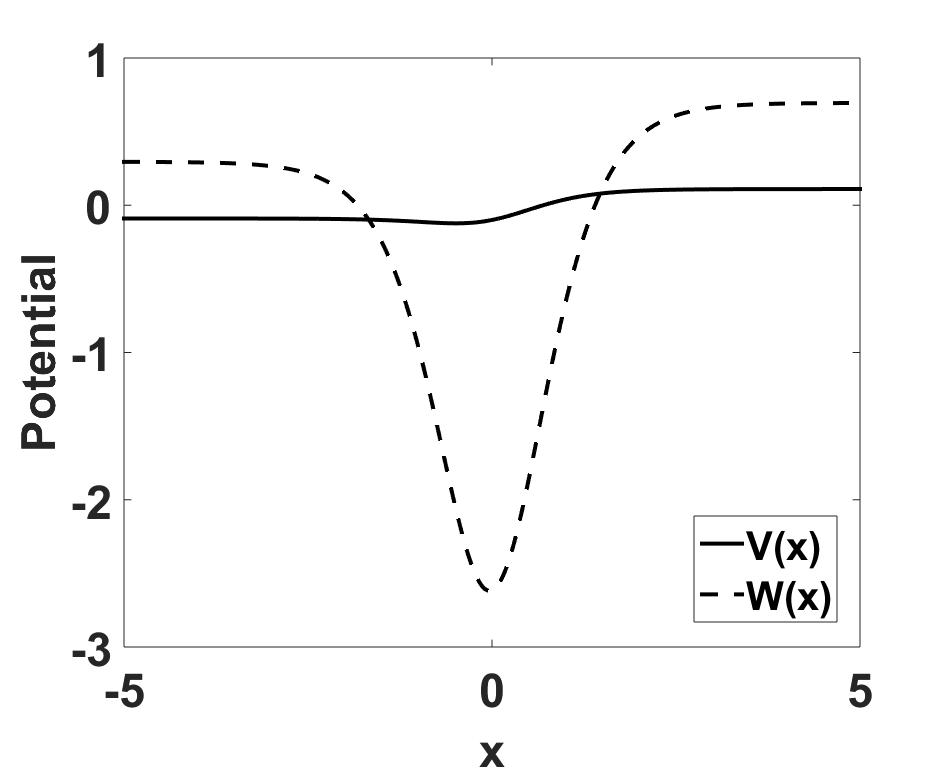}
\caption{}
\end{subfigure}
\begin{subfigure}[b]{0.325\textwidth}
\includegraphics[width=1\textwidth]{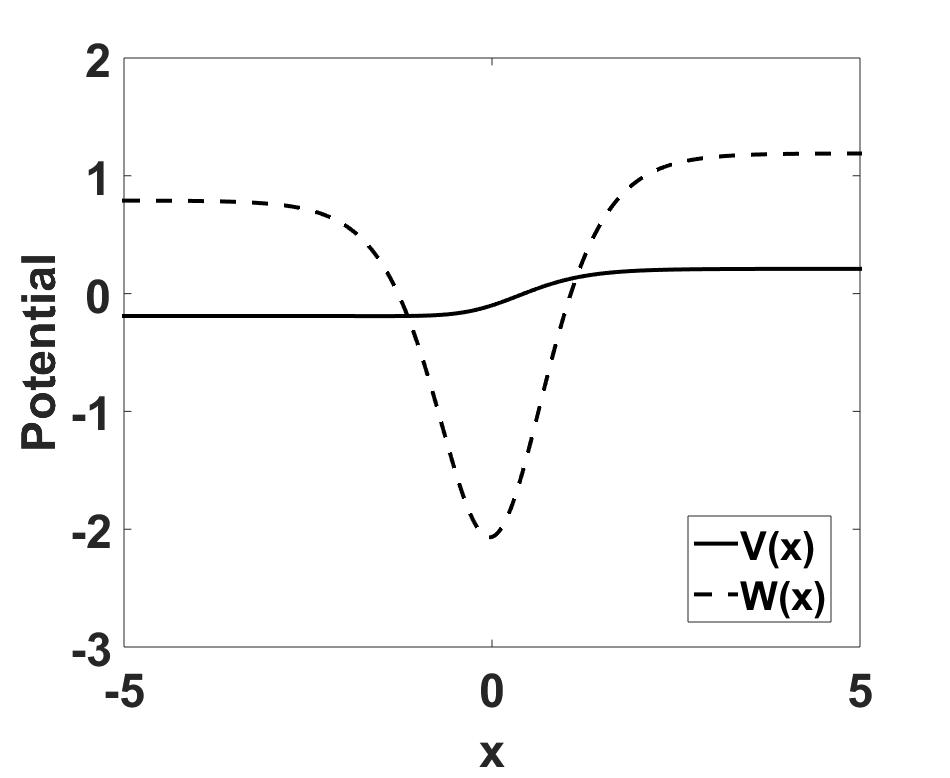}
\caption{}
\end{subfigure}
\begin{subfigure}[b]{0.325\textwidth}
\includegraphics[width=1\textwidth]{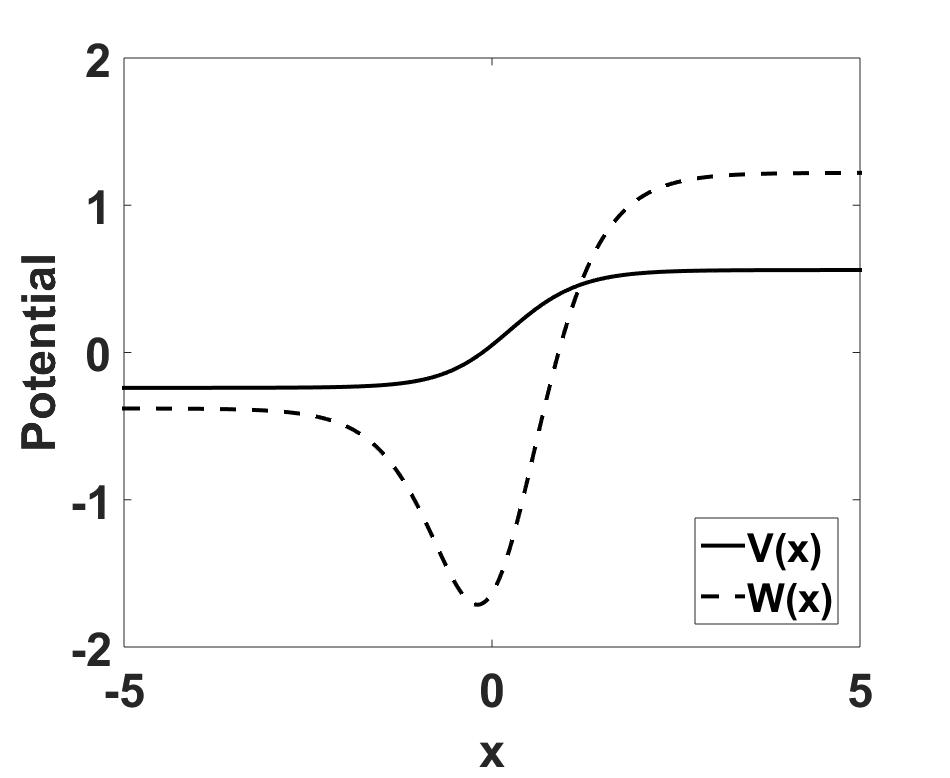}
\caption{}
\end{subfigure}
\begin{subfigure}[b]{0.325\textwidth}
\includegraphics[width=1\textwidth]{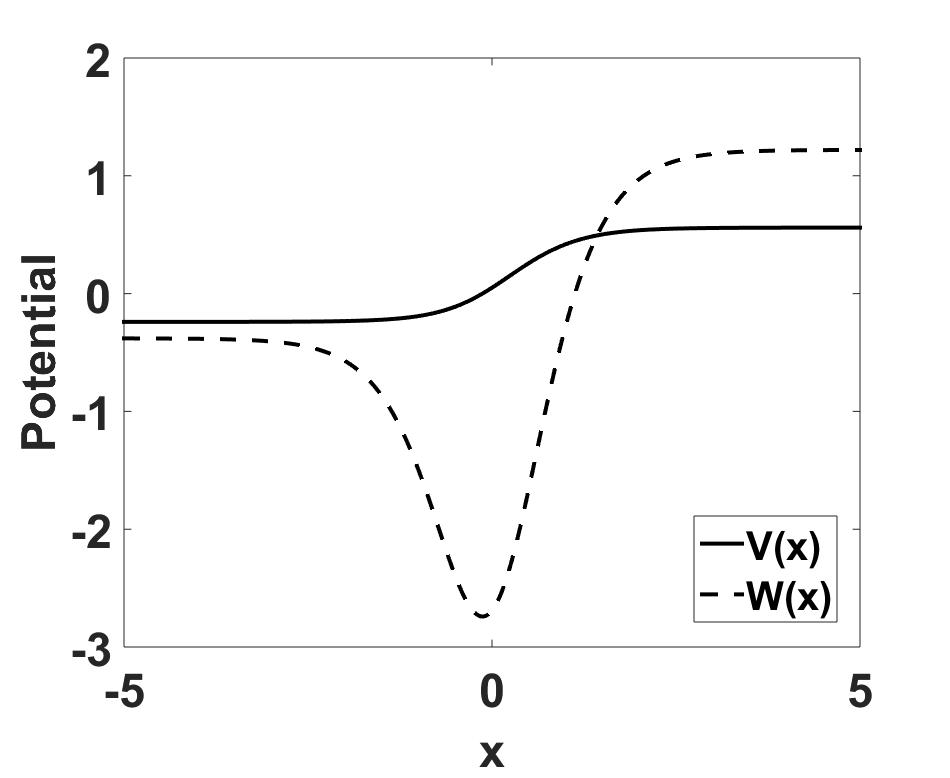}
\caption{}
\end{subfigure}
\begin{subfigure}[b]{0.325\textwidth}
\includegraphics[width=1\textwidth]{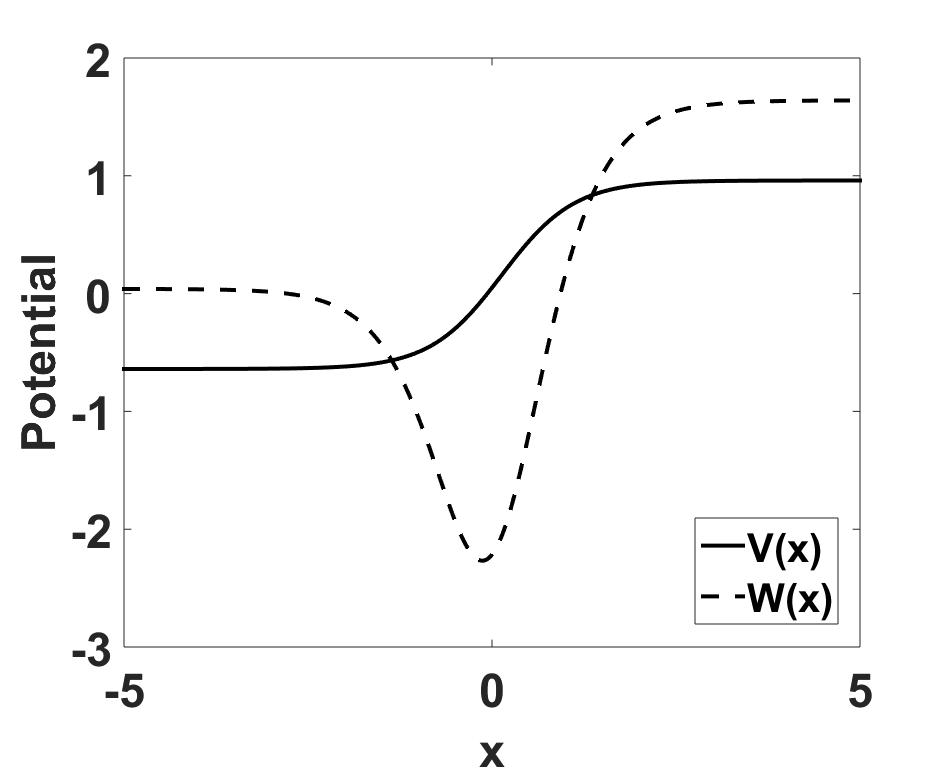}
\caption{}
\end{subfigure}
\begin{subfigure}[b]{0.325\textwidth}
\includegraphics[width=1\textwidth]{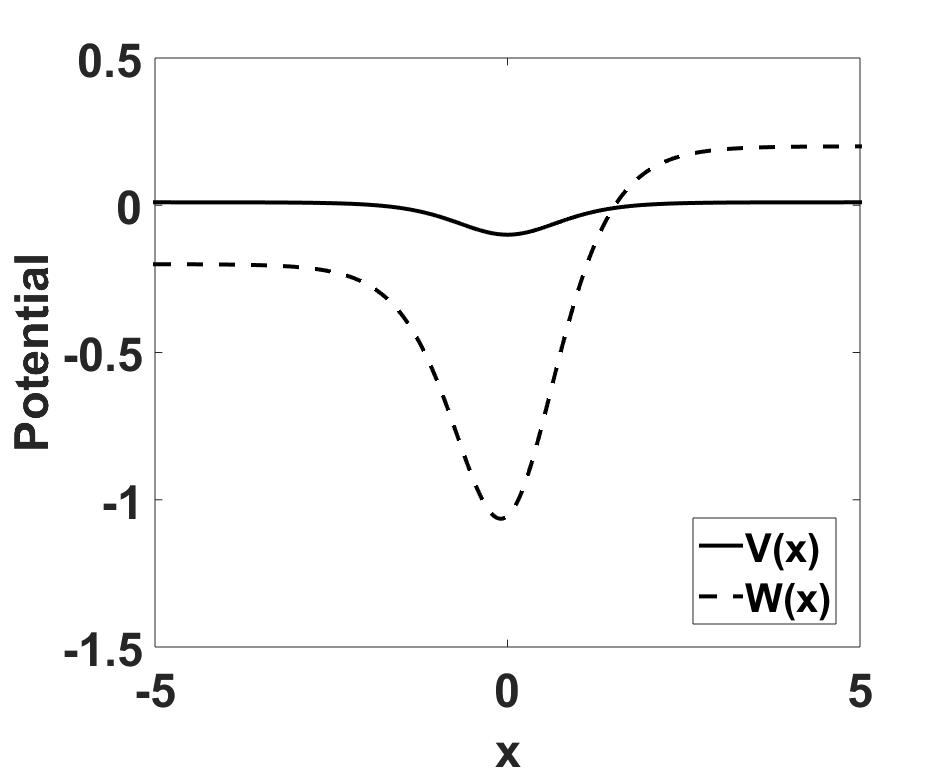}
\caption{}
\end{subfigure}
\begin{subfigure}[b]{0.325\textwidth}
\includegraphics[width=1\textwidth]{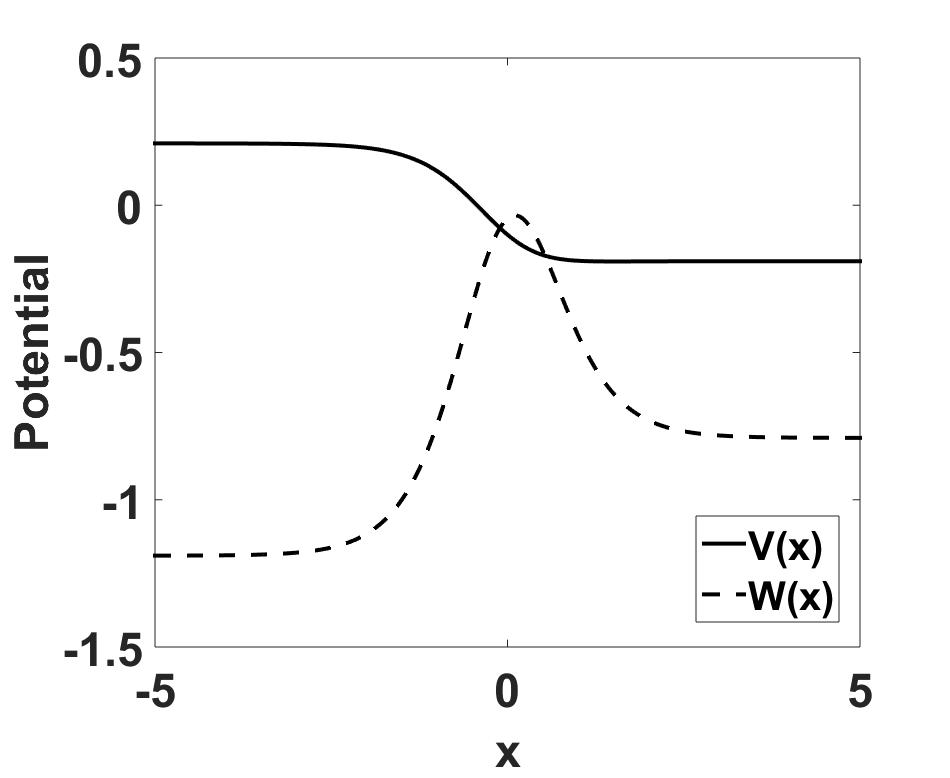}
\caption{}
\end{subfigure}
\caption{The nature of modified asymmetric Rosen-Morse potential, both $V(x)$ and $W(x)$. \textbf{(a)} $b=0.1$, $\alpha_{2}=-0.5$, $\beta_{2}=0.5$; \textbf{(b)} $b=0.1$, $\alpha_{2}=-0.5$, $\beta_{2}=1$; \textbf{(c)} $b=0.1$, $\alpha_{2}=-1$, $\beta_{2}=0.5$; \textbf{(d)} $b=0.4$, $\alpha_{2}=-0.5$, $\beta_{2}=0.5$; \textbf{(e)} $b=0.4$, $\alpha_{2}=-0.5$, $\beta_{2}=1$; \textbf{(f)} $b=0.4$, $\alpha_{2}=-1$, $\beta_{2}=0.5$;  \textbf{(g)} $b=0.1$, $\alpha_{2}=0$, $\beta_{2}=0.5$; and \textbf{(h)} $b=0.1$, $\alpha_{2}=1$, $\beta_{2}=0.5$}.
\label{fig2}
\end{figure*}
where $W_{1}=2\alpha_{2}-(a(a+1)+2\alpha_{1})\left( \frac{\beta_{2}}{\beta_{1}}\right)$, $a$ and $b$ are positive real values describing strength of the potential. If the potential is expressed as $V'(x)+ i  W'(x)$ such that 
\begin{subequations}\label{eq5}
\begin{align}
\label{eq5a} V'(x)= & \left(-a(a+1)+i W_1\right)\sech^{2}(x)+\frac{b^2}{\alpha_1} \\
\label{eq5b} W'(x)=& \left(1+i\left(\frac{\alpha_2}{\alpha_1}\right)\right)2b\tanh(x)+\alpha_2\left(\frac{b^2}{\alpha_1^2}-1\right),
\end{align}
\end{subequations}
then the modification given is essentially adding an imaginary term to $V(x)$ and $W(x)$ in Eq. (\ref{eq3}). In this sense, the potential we consider is one modification of Rosen-Morse potential.  Since the physical model (\ref{eq1}) contains many system parameters, for our convenience we fix the values as $a=0.1$ and $\alpha_{1}=\beta_{1}=1$ throughout. The existence of system parameters $\alpha_{2}$ and $\beta_{2}$ breaks the $\mathcal{PT}-$symmetry of the system. If the physical effects described by $\alpha_{2}$ and $\beta_{2}$ is very negligible then the system leads to $\mathcal{PT}-$symmetry. The qualitative nature of new potential (\ref{eq4}) for some parameter values is given in Fig. \ref{fig2}. Unlike the original Rosen-Morse potential, here both $V(x)$ and $W(x)$ are non-vanishing asymptotically.

We have taken values for parameters in such a way that the dependence of parameter is visible.  Initially we fix the value for $b$ in Figs. \ref{fig2}(a)-\ref{fig2}(c) and varies $\alpha_{2}$ and $\beta_{2}$ wherein the depth of the potential $W(x)$ increases significantly. Then we change the value of $b$ for the same variations of $\alpha_{2}$ and $\beta_{2}$ in Figs. \ref{fig2}(d)-\ref{fig2}(f). For Figs. \ref{fig2}(g) and \ref{fig2}(h) we take $\alpha_2$ $\geq 0$ where real part of potential is shifted. From Fig. \ref{fig2}(h) it is evident that imaginary part of potential is inverted for $\alpha_2 = 1$. The increase in value of $b$ is also contributing to the shape of the potential.  In addition, we also observe that the asymmetry of both the potentials increase with $b$.  These are the major factors that actually control the potential. In a physical system the values of $\alpha_{2}$ and $\beta_{2}$ are important because these values are externally controlled to make slight modifications to the potentials. In other words, we can say that these parameters can be used to tune the system to reach a particular regime where we can achieve stable propagation of light pulses as demonstrated below.

\subsection{Stationary soliton solution and linear stability analysis}
Our investigation of stationary solution starts with self-focusing nonlinearity, $\sigma=1$. Now the system admits an exact solution for Eq. (\ref{eq2}) of the form
\begin{eqnarray}\label{eq6}
\phi(x)=\sqrt{\frac{a(a+1)+2\alpha_{1}}{\beta_{1}}} \sech(x) e^{\frac{ibx}{\alpha_{1}}}.
\end{eqnarray}
For this particular solution, we impose the condition $\mu=\alpha_{1}$ for which the above is true.
\begin{figure}[h!]
\begin{subfigure}[b]{0.49\textwidth}
\includegraphics[width=1\textwidth]{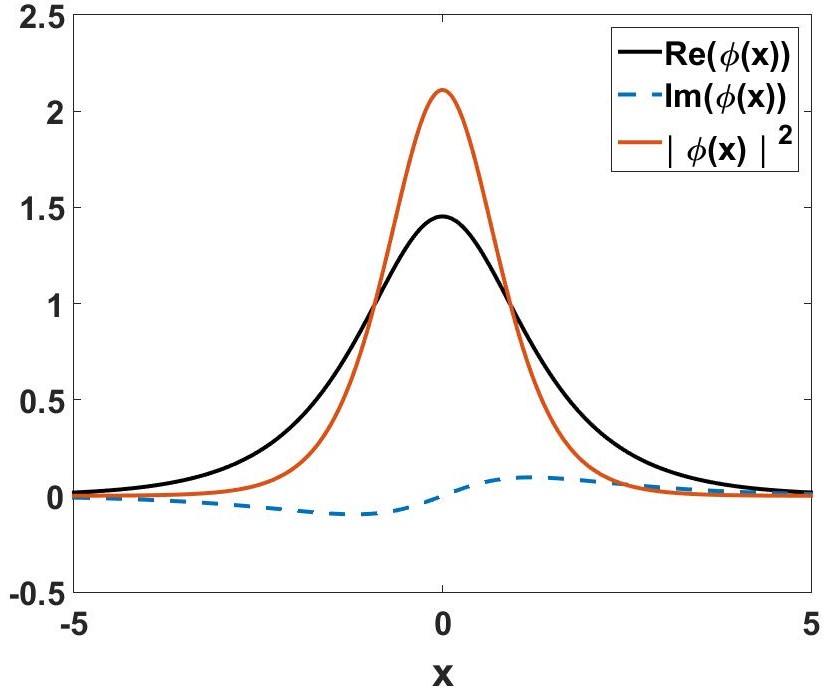}
\caption{}
\end{subfigure}
\begin{subfigure}[b]{0.49\textwidth}
\includegraphics[width=1\textwidth]{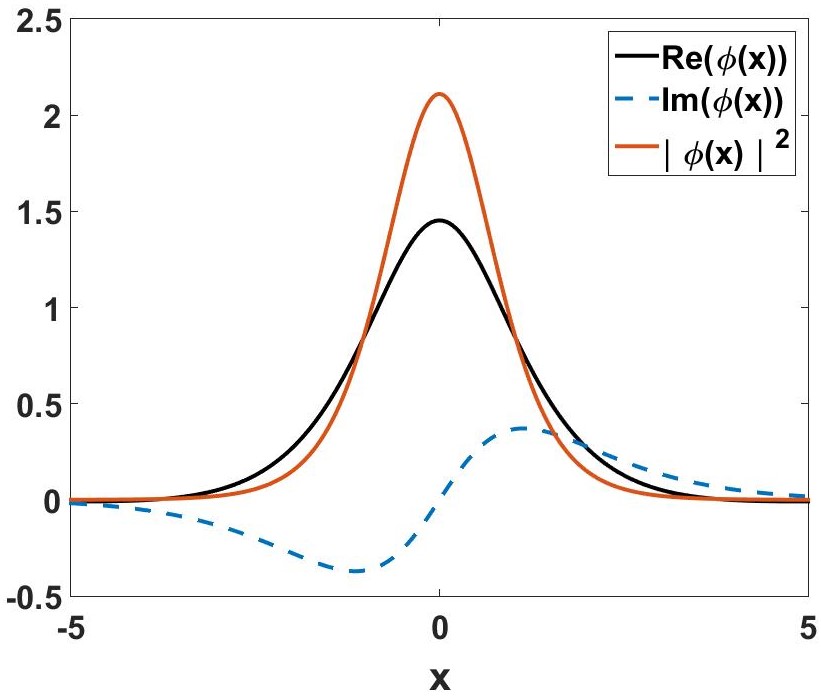}
\caption{}
\end{subfigure}
\caption{(Color online) Plot for stationary solution (\ref{eq6}) showing real part, imaginary part and $|\phi(x)|^2$ for \textbf{(a)} $ b=0.1$ and \textbf{(b)} $b=0.4.$}
\label{fig3}
\end{figure}
The profile of stationary soliton solution of Eq. (\ref{eq2}) is shown in Fig. \ref{fig3}. When we increase the parameter $b$ from $0.1$ to $0.4$, there is no change in total intensity of stationary solution.  But we can observe an increase in amplitude of imaginary part and decrease in the width of real part of stationary solution. 

The stability of this solution is evaluated by adding infinitesimal perturbation with the stationary solution. If such a perturbation leads to deviation from the original solution, then the solution is unstable. The perturbed solution \cite{24} is given by•••••••••••••••••••••••••
\begin{eqnarray}
\label{eq7}
\Psi(x,z) = (\phi(x)+\epsilon [f(x)e^{\lambda z}+g^{*}(x)e^{\lambda^{*} z}])e^{i\mu z},
\end{eqnarray}
where $\epsilon \ll 1$, $f(x)$ and $g(x)$ are very small perturbation functions and $\lambda$ is the stability parameter. Substituting the perturbed solution (\ref{eq7}) in Eq. (\ref{eq1}) and linearizing with respect to $\epsilon$, one can obtain a set of equations for $f(x)$ and $g(x)$ which can be solved by forming a matrix of the system of equations. 

\begin{figure*}[ht!]
\centering
\begin{subfigure}[b]{0.33\textwidth}•••••••••••••••••••••••••
\includegraphics[width=1\textwidth]{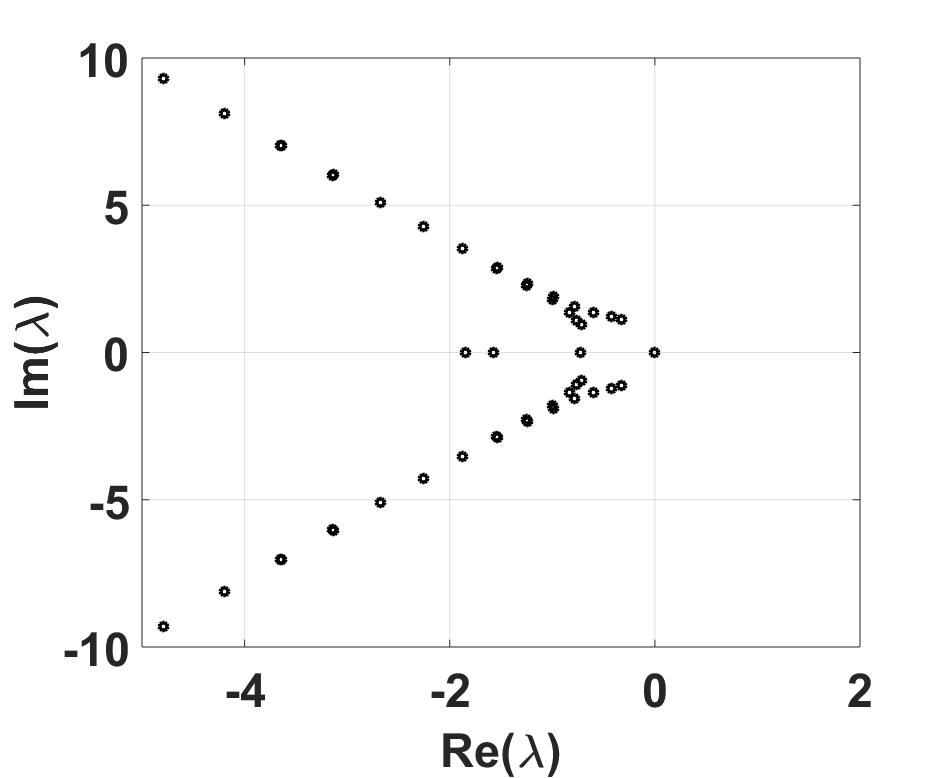}
\caption{}
\end{subfigure}
\begin{subfigure}[b]{0.33\textwidth}
\includegraphics[width=1\textwidth]{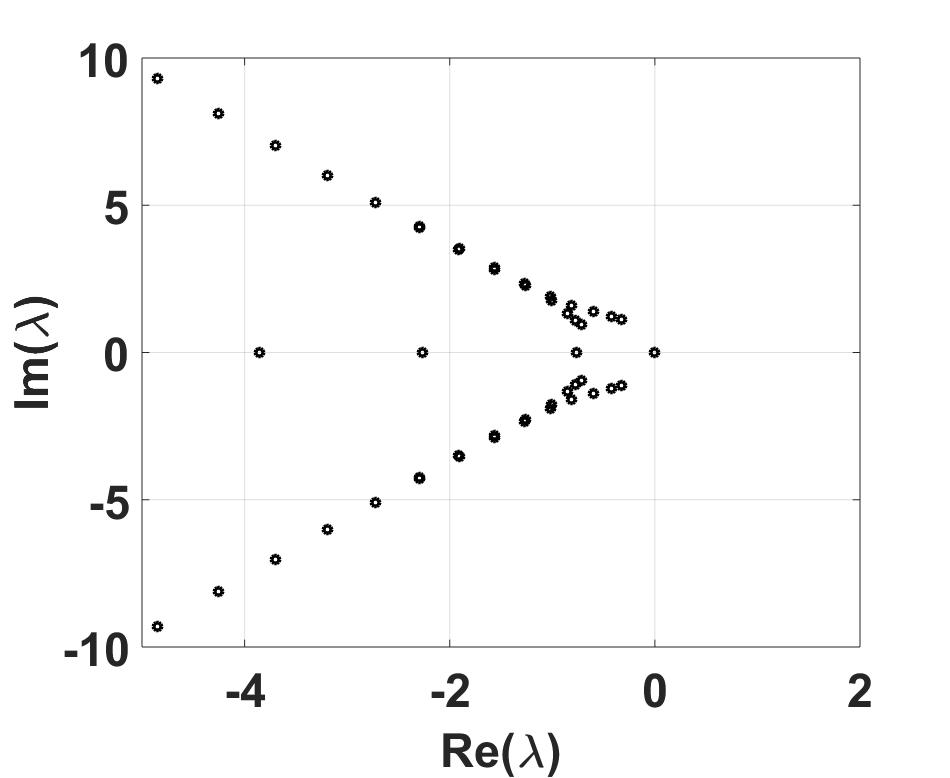}
\caption{}
\end{subfigure}
\begin{subfigure}[b]{0.32\textwidth}
\includegraphics[width=1\textwidth]{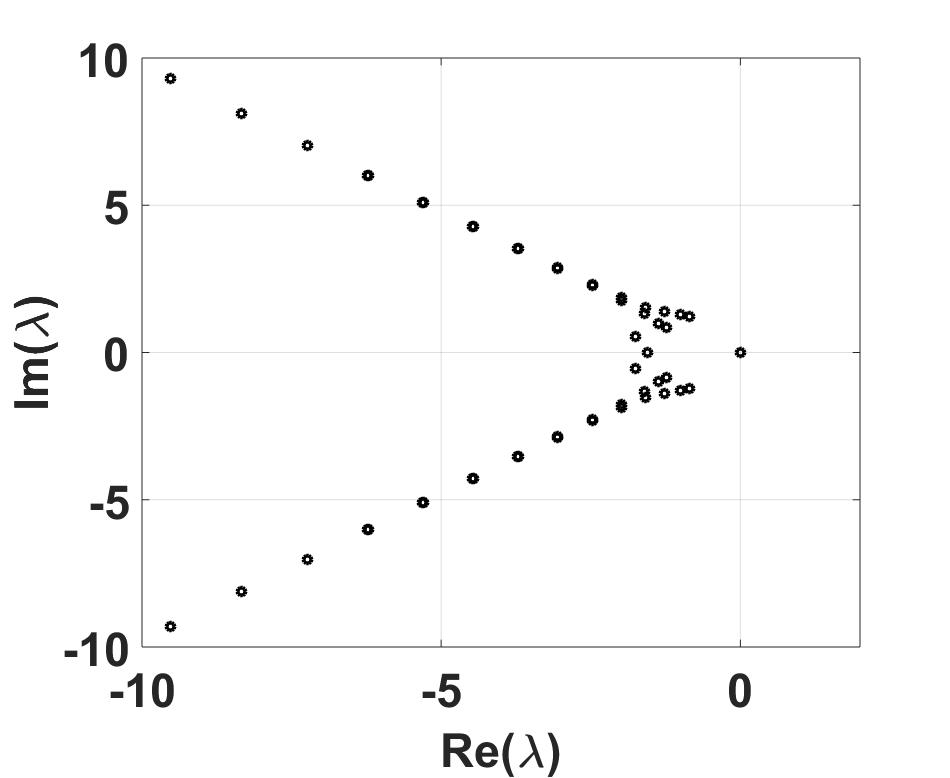}
\caption{}
\end{subfigure}
\begin{subfigure}[b]{0.33\textwidth}
\includegraphics[width=1\textwidth]{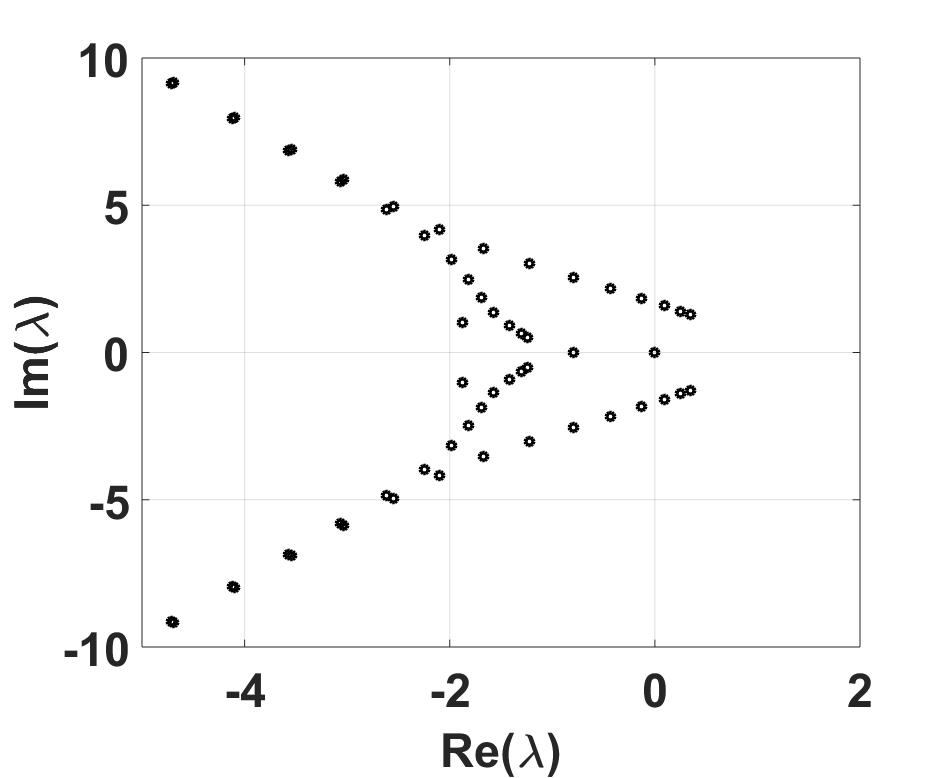}
\caption{}
\end{subfigure}
\begin{subfigure}[b]{0.33\textwidth}
\includegraphics[width=1\textwidth]{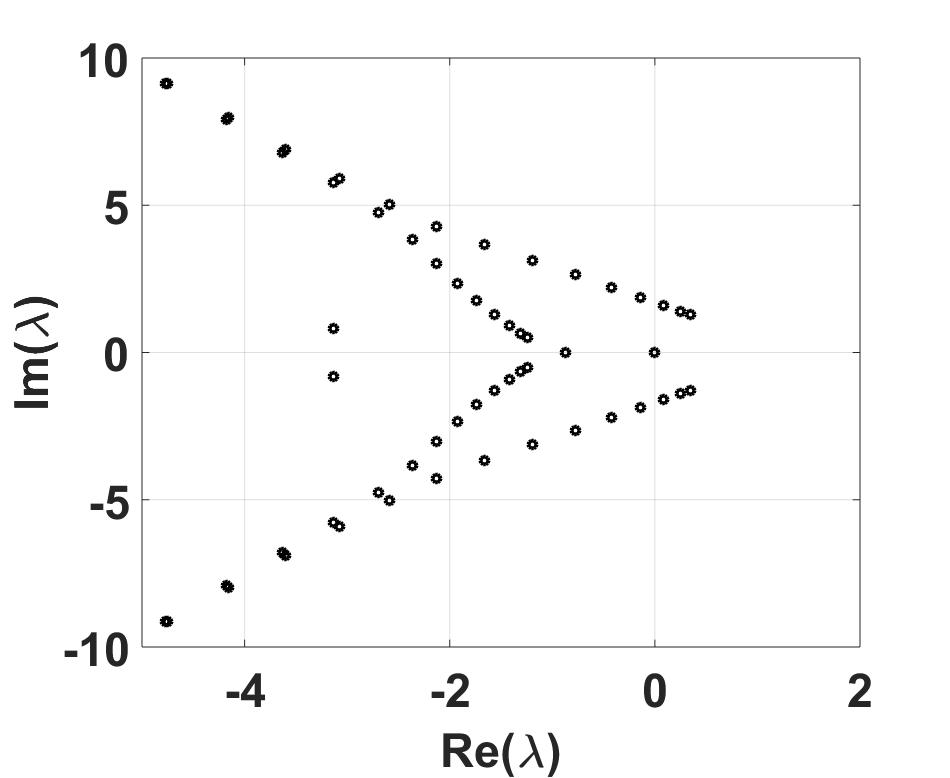}
\caption{}
\end{subfigure}
\begin{subfigure}[b]{0.32\textwidth}
\includegraphics[width=1\textwidth]{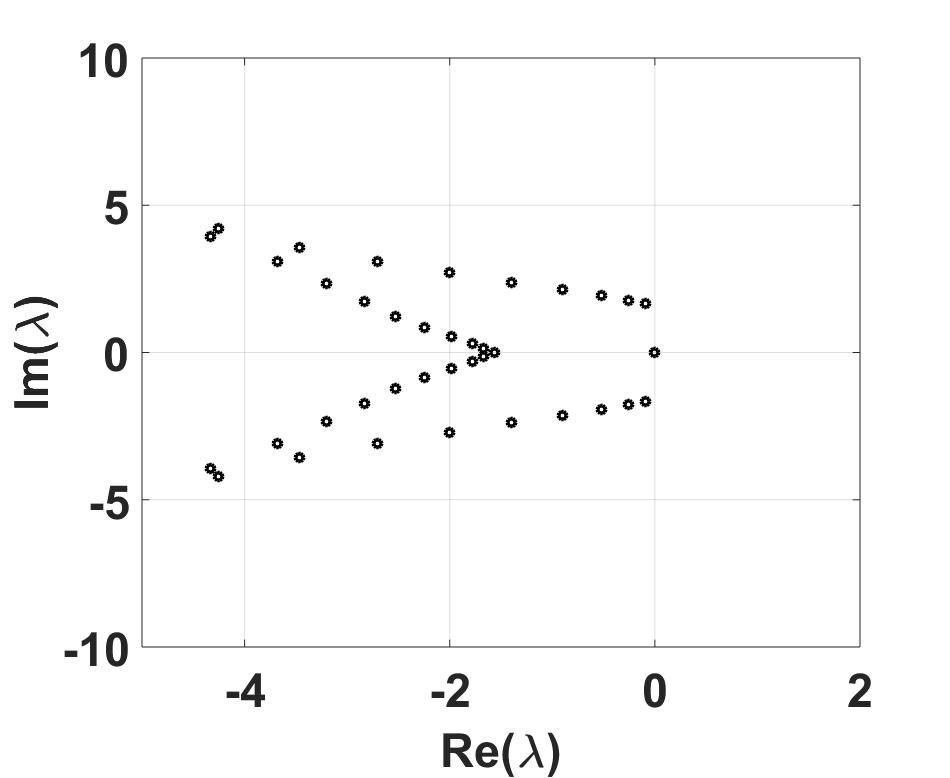}
\caption{}
\end{subfigure}
\begin{subfigure}[b]{0.33\textwidth}
\includegraphics[width=1\textwidth]{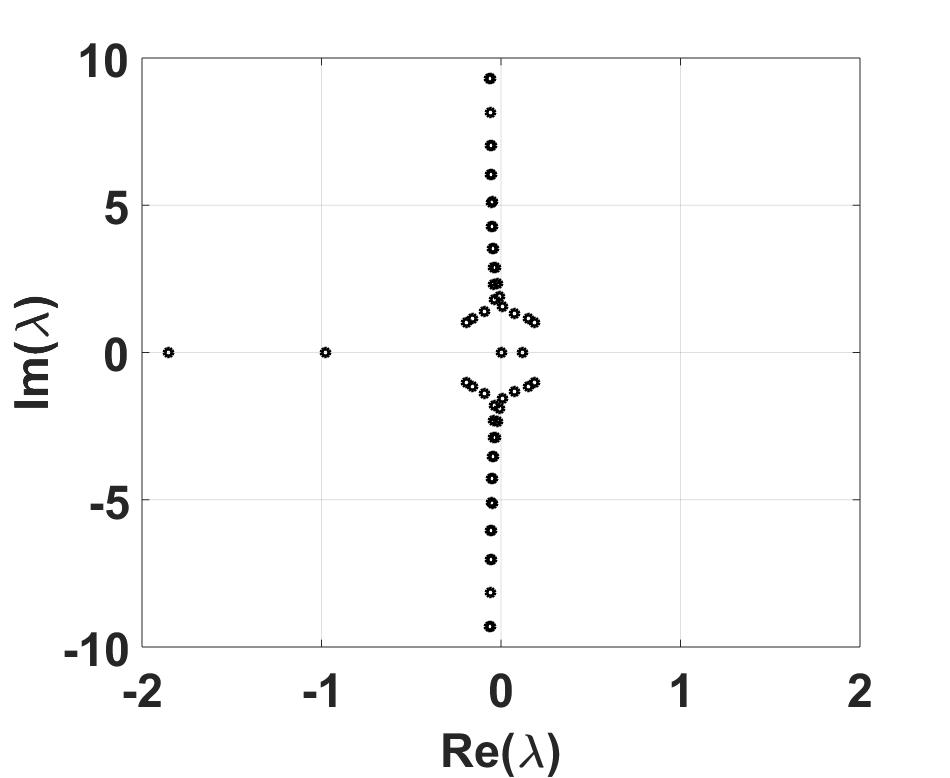}
\caption{}
\end{subfigure}
\begin{subfigure}[b]{0.33\textwidth}
\includegraphics[width=1\textwidth]{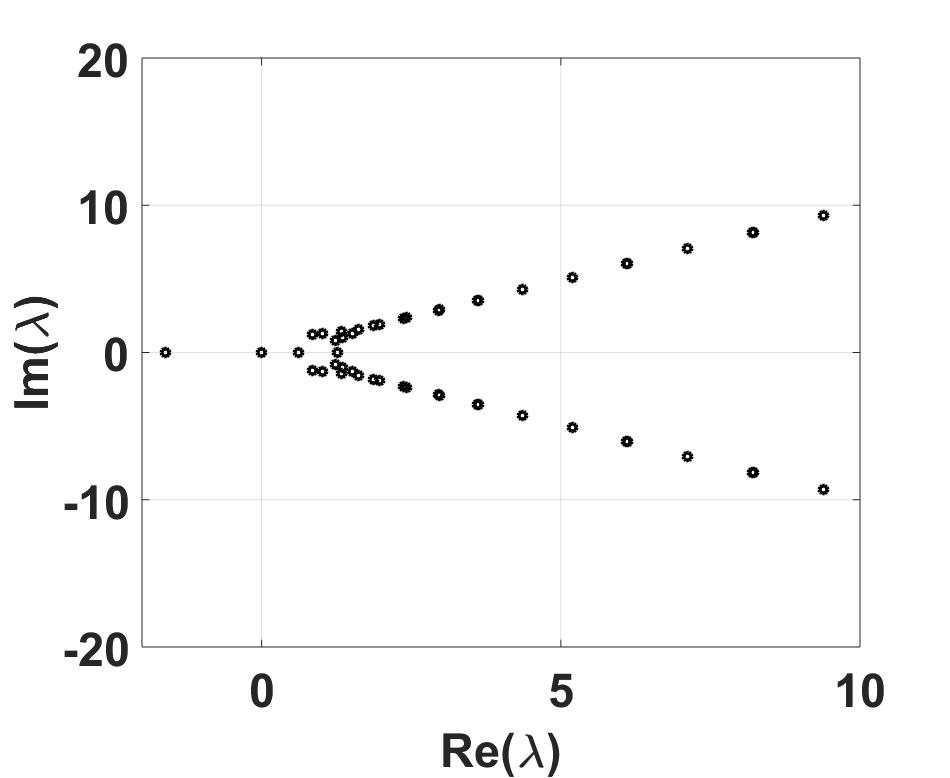}
\caption{}
\end{subfigure}
\caption{\textbf{(a)}$-$\textbf{(h)} Eigenvalue spectrum of stability analysis of Eq. (\ref{eq2}) for the same order of parameters as in Fig. \ref{fig2}}
\label{fig4}
\end{figure*}

The resulting linear eigenvalue problem is described by 
\begin{gather}
\begin{pmatrix} 
L_{0} &  L_{1}\\
 -L_{1}^{*} & -L_{0}^{*} 
\end{pmatrix}  
\begin{pmatrix}
f(x)\\
g(x)
\end{pmatrix}=-i\lambda
\begin{pmatrix}
f(x)\\
g(x)
\end{pmatrix},
\end{gather}
where $L_{0}=(\alpha_{1}+i\alpha_{2})\partial_{xx}+V(x)+iW(x)+2(\beta_{1}+i\beta_{2})|\phi|^{2}-\mu$ and $L_{1}=(\beta_{1}+i\beta_{2})\phi^{2}$.
This eigenvalue equation can be solved using Fourier collocation method \cite{25}.  Stable solutions are present only if the eigenvalues have a negative real part or purely imaginary. This means that when the system is perturbed from a singular point, the perturbation will eventually decay for negative real part of the eigenvalues. The stationary solution with purely imaginary eigenvalues oscillates around the singular point for small perturbation.  In case of positive eigenvalues, the terms in perturbation diverge and stability cannot be achieved even for infinitesimal perturbation. Figure \ref{fig4} illustrates the eigenvalues for different parameter values.

Thus we have obtained the parameter values for stable and unstable modes using linear stability analysis.  The stable regions are small ranges in different parameters.  Numerically it is found that all soliton solutions for our considered system are stable only for $\alpha_{2}<0$.  In Fig. \ref{fig4}, we have plotted the real and imaginary parts of eigenvalues $\lambda$ for different parameters.  In the panel Figs. \ref{fig4}(a)$-$\ref{fig4}(d), the eigenvalues are shown to be complex with negative real parts, implying that the perturbed solution oscillates and eventually dies off.  On the other hand, Figs. \ref{fig4}(e)$-$\ref{fig4}(h) show eigenvalues with positive real part showing that the solutions are unstable.

\begin{figure*}[!ht]
\centering
\begin{subfigure}[b]{0.32\textwidth}
\includegraphics[width=1\textwidth]{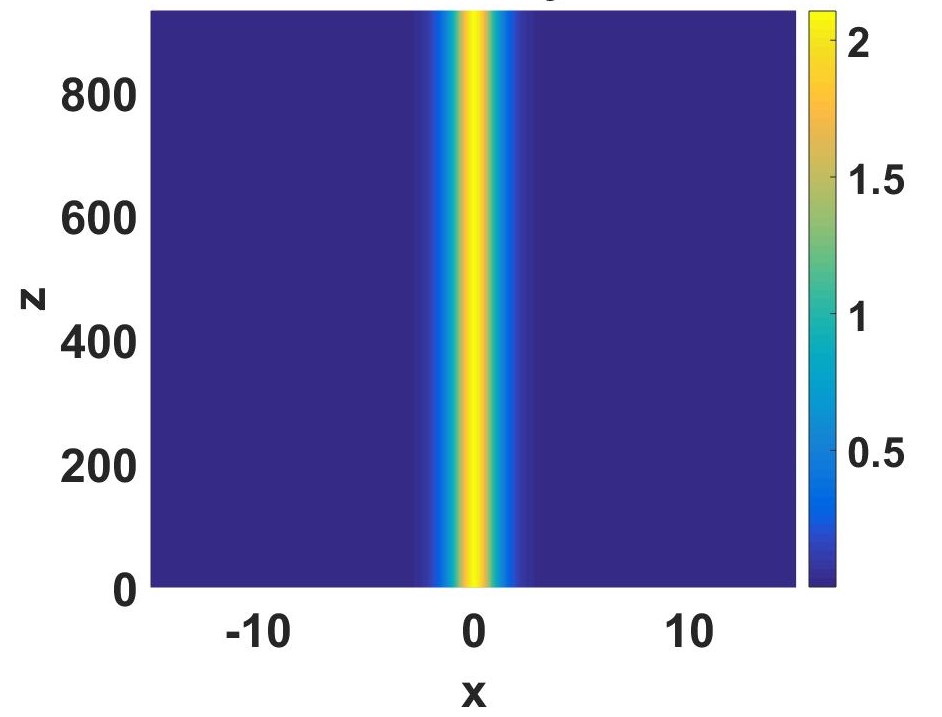}
\caption{}
\end{subfigure}
\begin{subfigure}[b]{0.32\textwidth}
\includegraphics[width=1\textwidth]{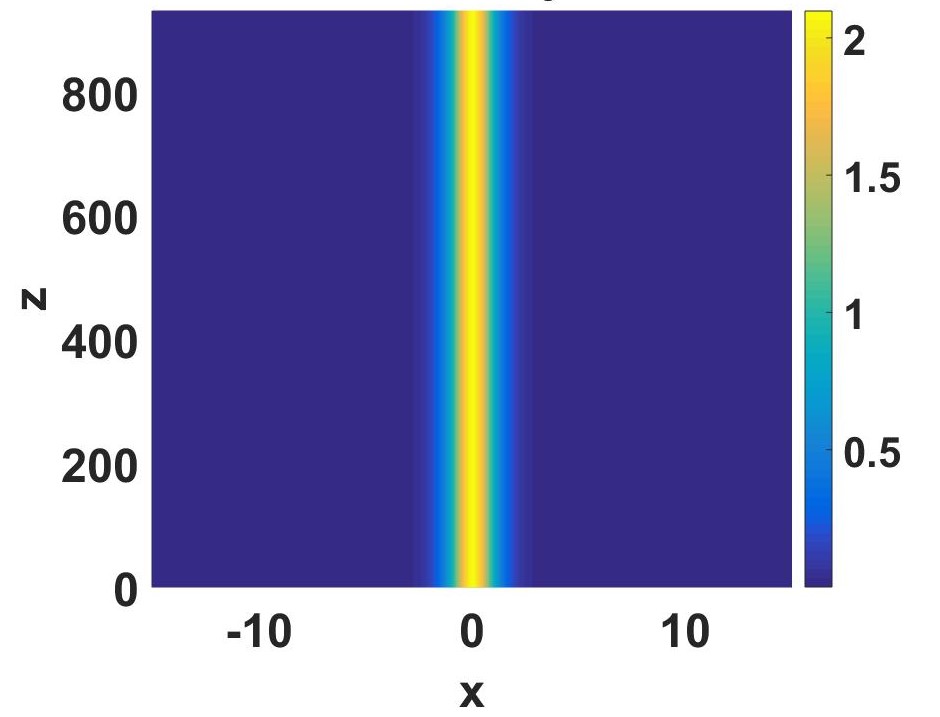}
\caption{}
\end{subfigure}
\begin{subfigure}[b]{0.32\textwidth}
\includegraphics[width=1\textwidth]{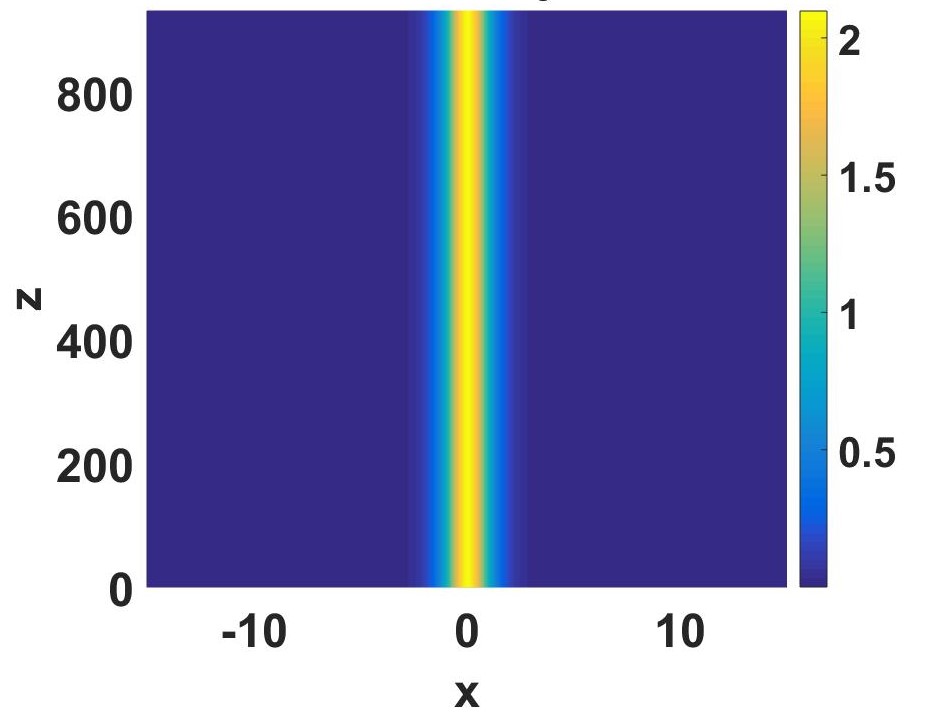}
\caption{}
\end{subfigure}
\begin{subfigure}[b]{0.32\textwidth}
\includegraphics[width=1\textwidth]{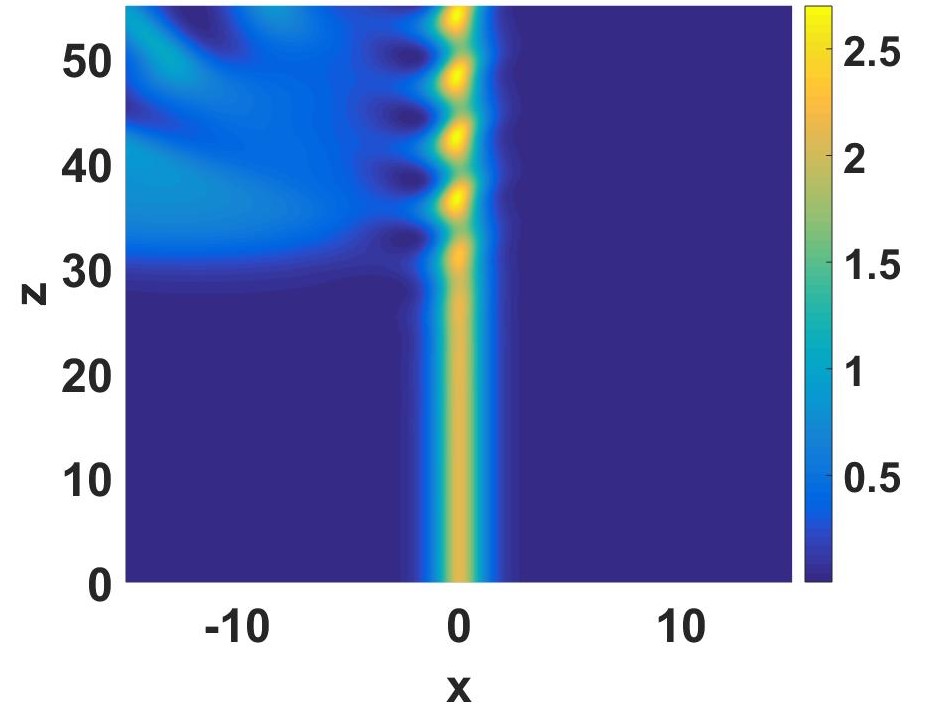}
\caption{}
\end{subfigure}
\begin{subfigure}[b]{0.32\textwidth}
\includegraphics[width=1\textwidth]{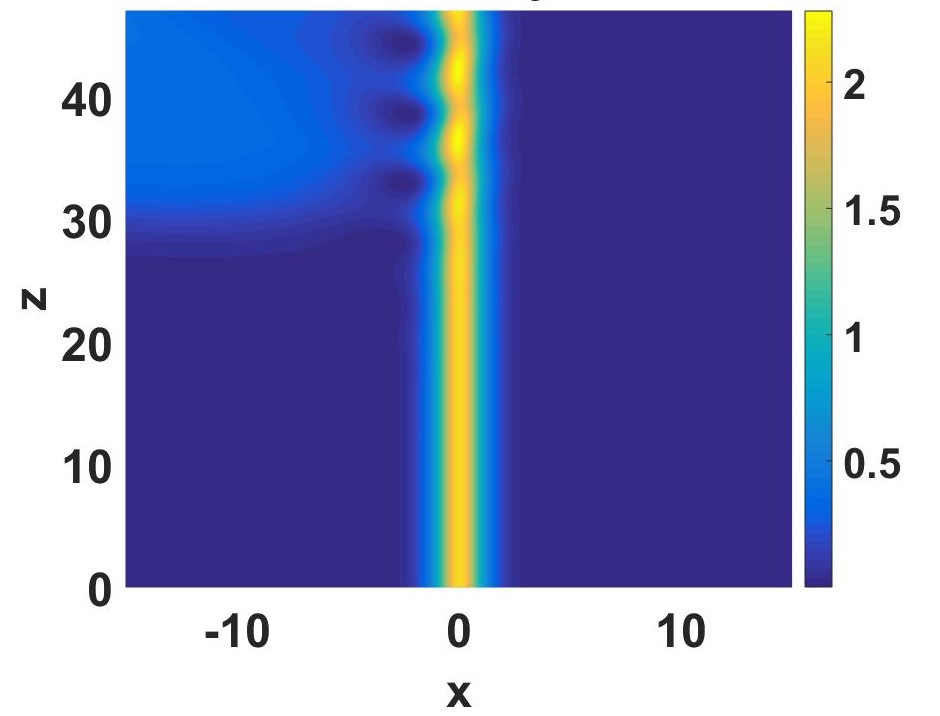}
\caption{}
\end{subfigure}
\begin{subfigure}[b]{0.32\textwidth}
\includegraphics[width=1\textwidth]{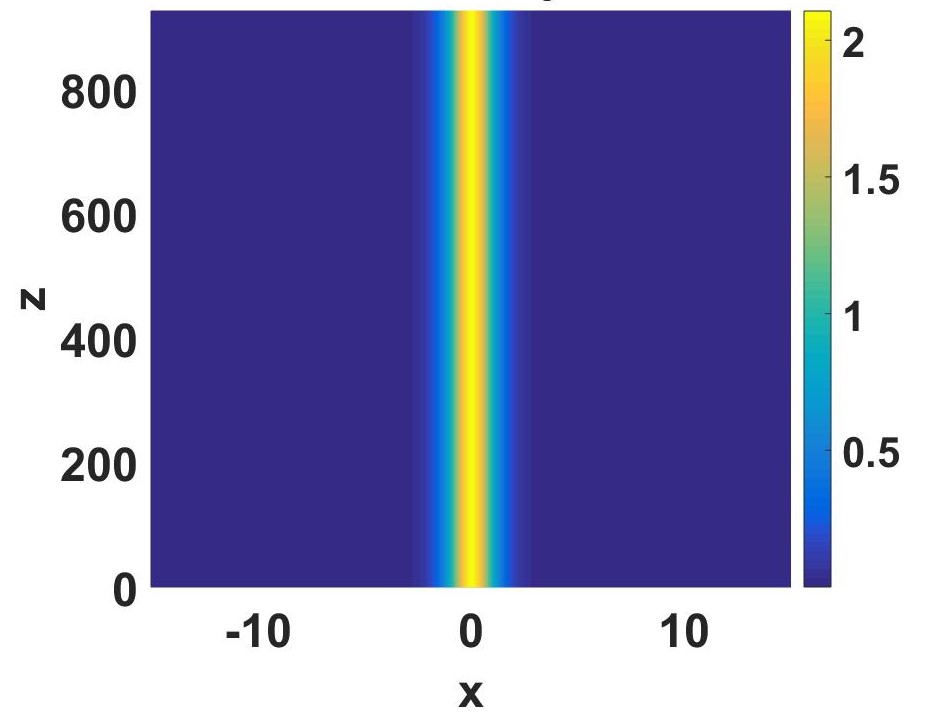}
\caption{}
\end{subfigure}
\begin{subfigure}[b]{0.33\textwidth}
\includegraphics[width=1\textwidth]{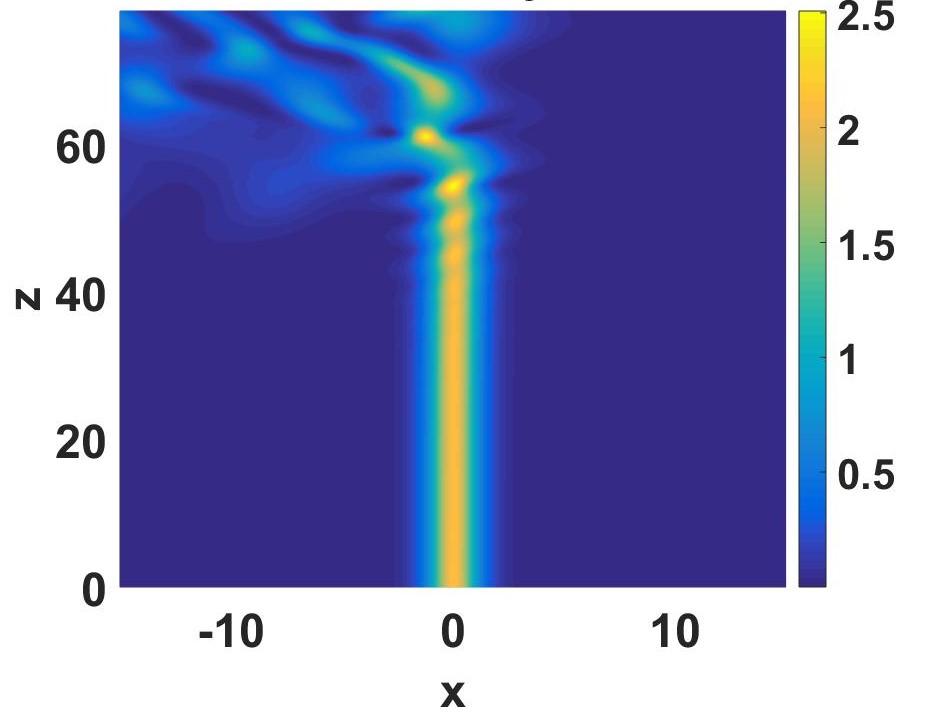}
\caption{}
\end{subfigure}
\begin{subfigure}[b]{0.33\textwidth}
\includegraphics[width=1\textwidth]{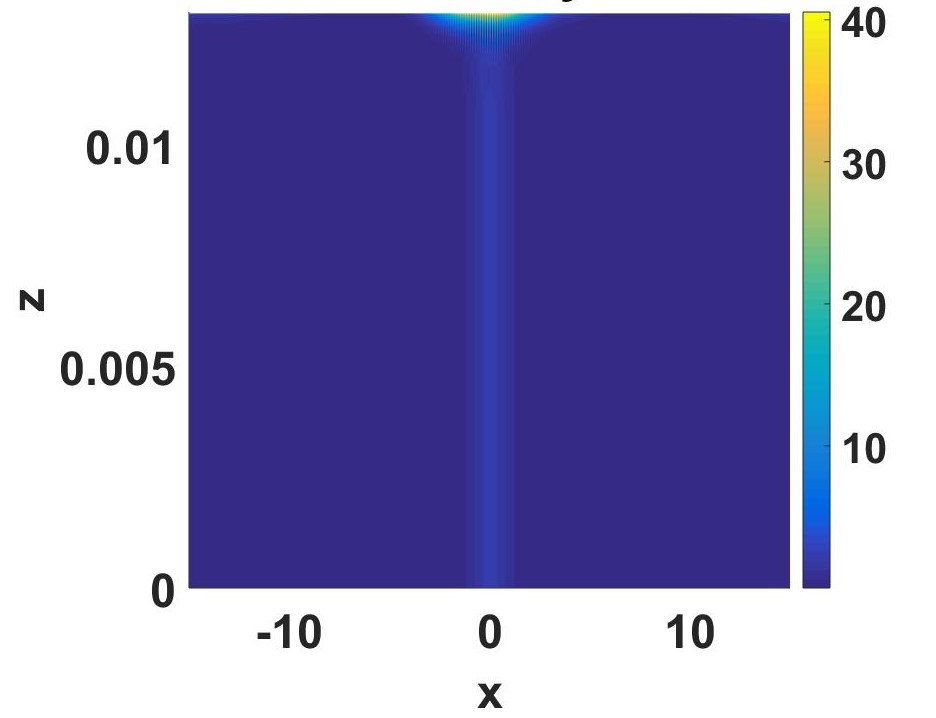}
\caption{}
\end{subfigure}
\caption{(Color online) \textbf{(a)}$-$\textbf{(h)} Intensity profile for stable evolution of soliton along the $z$ direction for the same order of parameters as in Fig. \ref{fig2}}
\label{fig5}
\end{figure*}

\begin{figure*}[!ht]
\begin{subfigure}[b]{0.49\textwidth}
\includegraphics[width=1\textwidth]{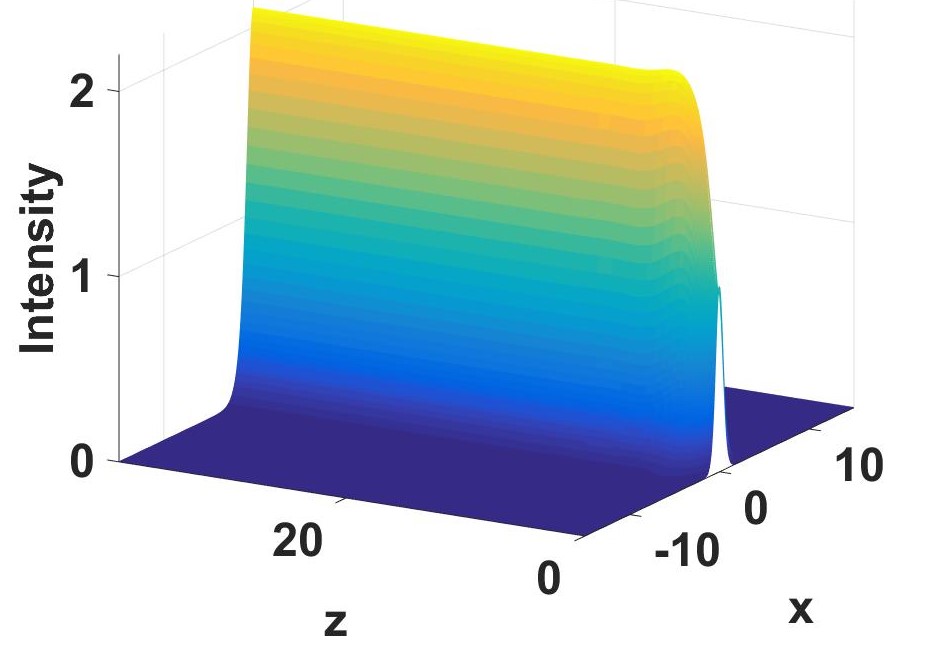}
\caption{}
\end{subfigure}
\begin{subfigure}[b]{0.49\textwidth}
\includegraphics[width=1\textwidth]{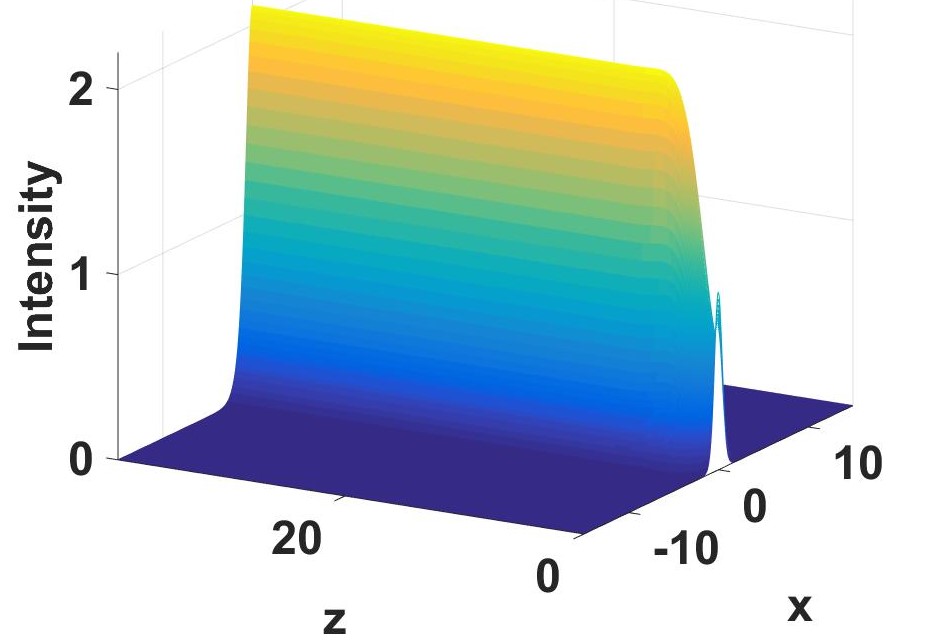}
\caption{}
\end{subfigure}
\caption{Intensity profile for different initial condition $\phi(x)=e^{\frac{ibx}{2}}\sech^3(x)$ of (\ref{eq2}) with parameters as \textbf{(a)} $b=0.1$, $\alpha_{2}=-0.5$, $\beta_{2}=1$ and \textbf{(b)} $b=0.1$, $\alpha_{2}=-1$, $\beta_{2}=0.5$.}
\label{fig5-1}
\end{figure*}

\begin{figure*}[!ht]
\begin{subfigure}[b]{0.49\textwidth}
\includegraphics[width=1\textwidth]{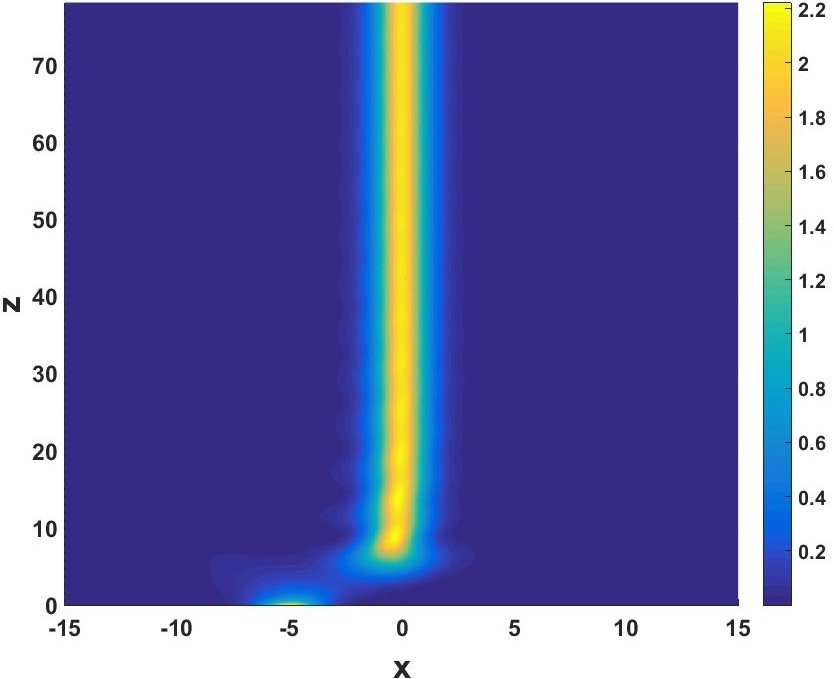}
\caption{}
\end{subfigure}
\begin{subfigure}[b]{0.49\textwidth}
\includegraphics[width=1\textwidth]{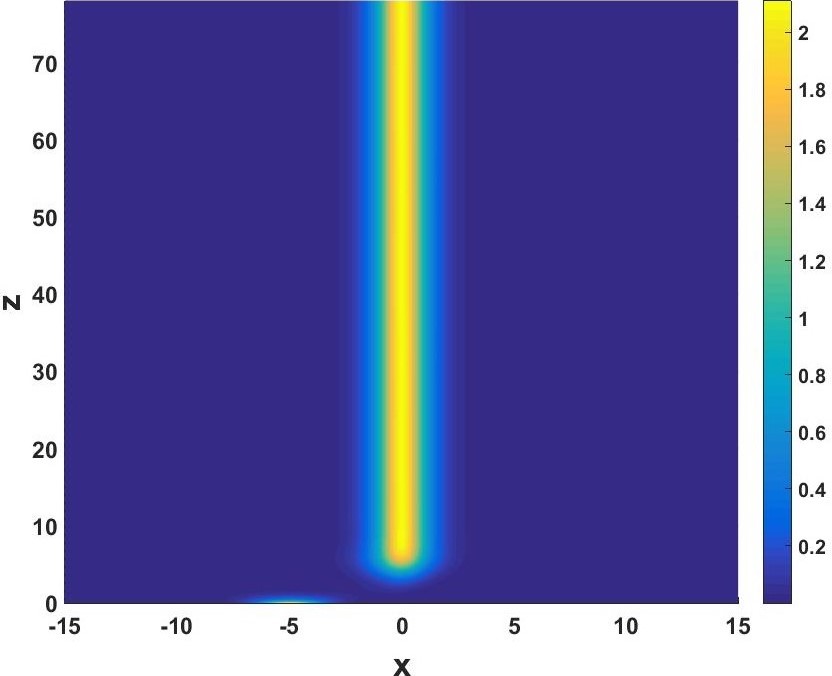}
\caption{}
\end{subfigure}
\caption{(Color online) Intensity profile for shift in position of soliton with \textbf{(a)} $b=0.1$, $\alpha_2=-0.2$, $\beta_2=0.5$ and \textbf{(b)} $b=0.1$, $\alpha_2=-1$, $\beta_2=0.5$.}
\label{fig5_1}
\end{figure*}

We follow the convention that the soliton propagates in $z-$ direction.  Figure \ref{fig5} represents intensity plot of solitons for the same parameter values as in Fig. \ref{fig4}. We clearly observe that solitons are stable in Figs. \ref{fig5}(a)-\ref{fig5}(d).  On the other hand, Figs. \ref{fig5}(e)-\ref{fig5}(f) show that solitons propagate such that the amplitudes oscillate with their location being unstable, as expected from the stability analysis.  There also exist some cases where the amplitude of solitons oscillates during propagation without affecting their stability, which is one of the important properties of dissipative solitons \cite{1}. In Fig. \ref{fig5}(g) the soliton is unstable, deviates from its path and dissipates. In Fig. \ref{fig5}(h) as $\alpha_{2}$ increases, the soliton is not able to propagate even a little distance, where it scatters due to $\alpha_2 > 0$. For stable soliton solutions, the maximum intensity is at $x=0$ plane and reduces to zero when $x\rightarrow\pm \infty$. 

To determine the stability for a different initial condition we have used another initial condition $\phi(x)=e^{\frac{ibx}{2}}\sech^3(x)$. The initial condition self-organizes into the original soliton profile. This solution is stable and propagates without any dissipation as shown in Figs. \ref{fig5-1}(a) and \ref{fig5-1}(b) for different values of $\alpha_2$ and $\beta_2$. Since the shape of soliton is $\sech-$like, we have used a similar function for the initial condition. We have also observed that shifting the initial position of stable soliton from centre of the potential does not affect the stability, as shown in Figs. \ref{fig5_1}(a) and \ref{fig5_1}(b). Here the soliton reorients into the path $x=0$. Thus, we find that the complex asymmetric potential accommodates dissipative soliton solutions. 

\subsection{Energy flow for exact soliton solution}
\begin{figure}[h!]
\begin{subfigure}[b]{0.49\textwidth}
\includegraphics[width=1\textwidth]{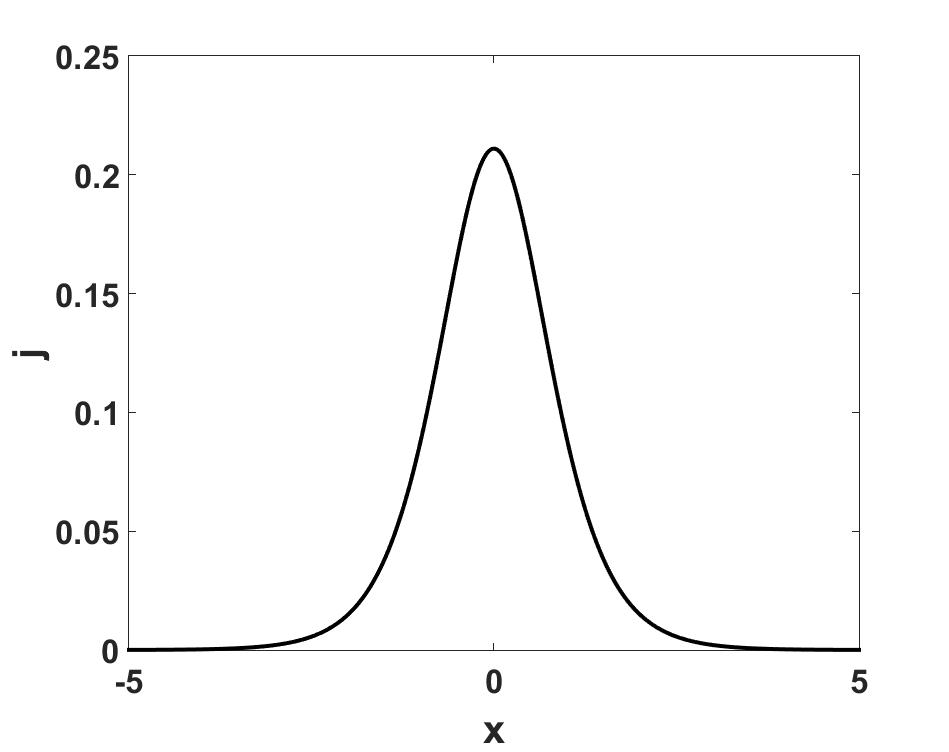}
\caption{}
\end{subfigure}
\begin{subfigure}[b]{0.49\textwidth}
\includegraphics[width=1\textwidth]{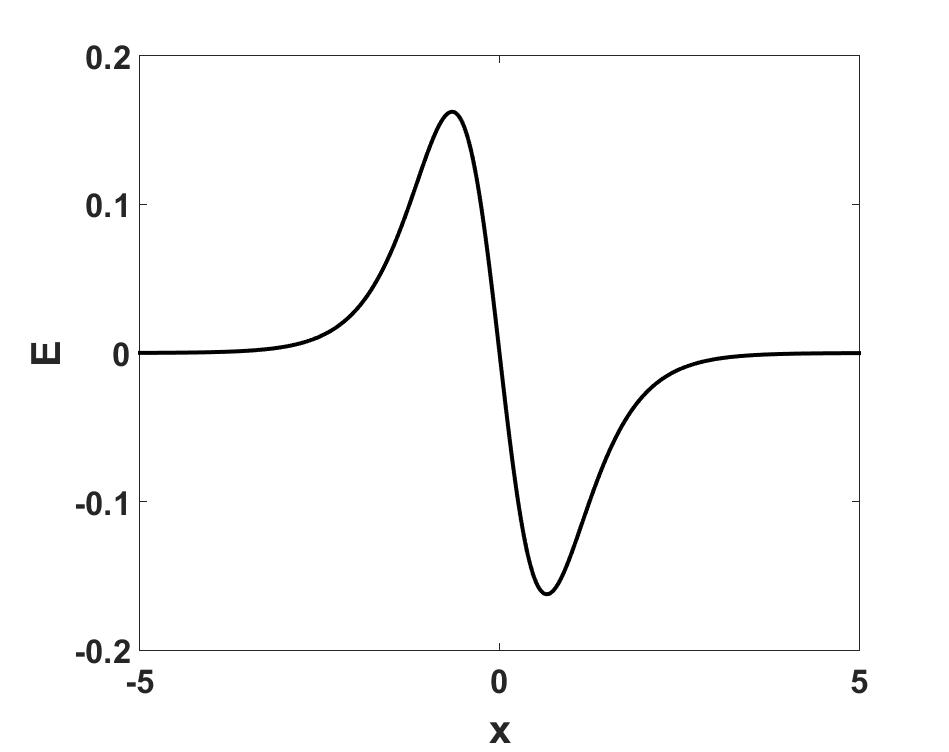}
\caption{}
\end{subfigure}
\caption{\textbf{(a)} Energy flux $(j)$ with respect to spatial coordinate for self-focusing mode with $b=0.1$ and \textbf{(b)} The gain or loss distribution of energy $(E)$ with respect to spatial coordinate for $b=0.1$}
\label{fig6}
\end{figure}

Equation of continuity for energy flow (where the energy is not conserved) of CGL Eq. (\ref{eq1}) gives us the relation between energy density ($\rho = |\Psi|^2$) and energy flux $(j)$ as \cite{21} 
\begin{eqnarray}\label{eq10}
E = \frac{\partial\rho}{\partial z}+\frac{\partial j}{\partial x},
\end{eqnarray}
where $E$ determines the gain or loss distribution of energy and the energy flux is described by $j=\frac{i}{2}(\phi\phi_{x}^{*}-\phi_{x}\phi^{*})$.  If the system is conservative then $E=0$ which means the loss and gain in the system is balanced. When the loss and gain of a system are not balanced, energy will flow from one region to another region.  For the system of our interest with $\sigma=1$, we find that
\begin{subequations}
\begin{align}\label{eq11}
E=& \frac{-2b}{\alpha_{1} \beta_{1}}(a(a+1)+2\alpha_{1})\text{sech}^{2}(x)\text{tanh}(x), \\
j=& \frac{b\left(a(a+1)+2\alpha_1\right)}{\alpha_1\beta_1}\sech^2(x).
\end{align}
\end{subequations}
This result implies that energy flux is similar to the intensity profile of stable soliton, with maximum flux through $x=0$, as shown in Fig. \ref{fig6}(a).  
The energy flux of course depends on the value of $b$ since $a, \alpha_{1}$  and $\beta_{1}$ are fixed. The direction and quantity of energy flow are illustrated in Fig. \ref{fig6}(b). Here we observe that the light pulse gains energy in $x<0$ region and looses energy in $x>0$ region.  In other words, the energy flow is effected from left to right.  An important observation from Eq. (\ref{eq11}) is that the energy flow due to unbalanced loss and gain does not depend on $\alpha_{2}$ and $\beta_{2}$.  In other words, the energy flow does not depend on spectral filtering and nonlinear gain/loss of the optical system under consideration.  
\section{Self-defocusing nonlinear mode}
\subsection{Nature of potential}
\begin{figure}[h!]
\begin{subfigure}[b]{0.33\textwidth}
\includegraphics[width=1\textwidth]{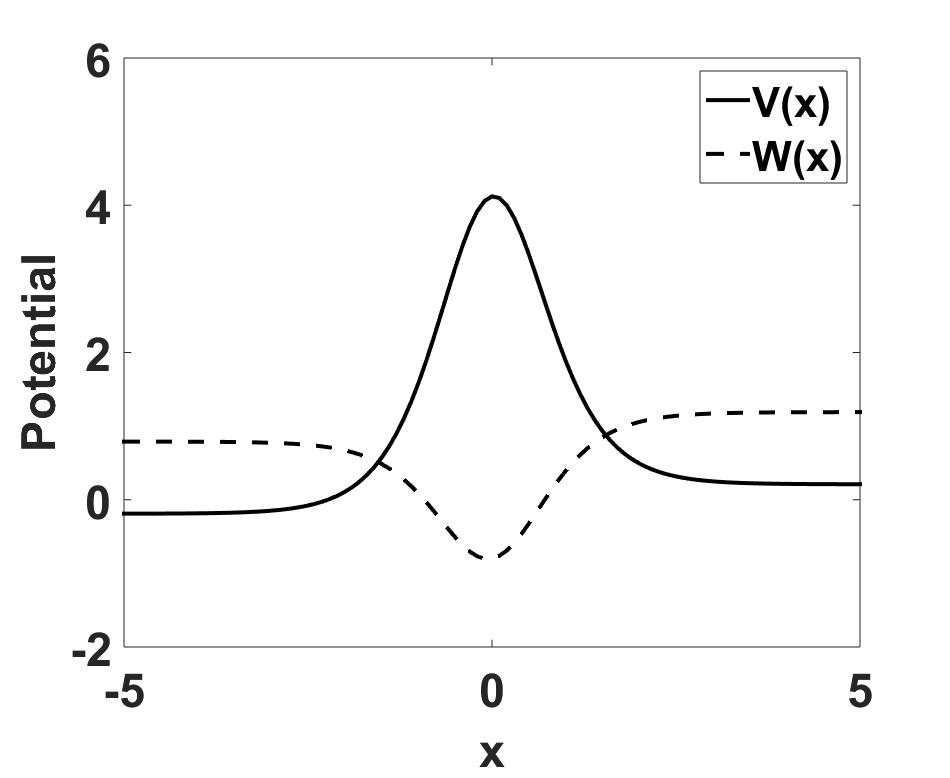}
\caption{}
\end{subfigure}
\begin{subfigure}[b]{0.33\textwidth}
\includegraphics[width=1\textwidth]{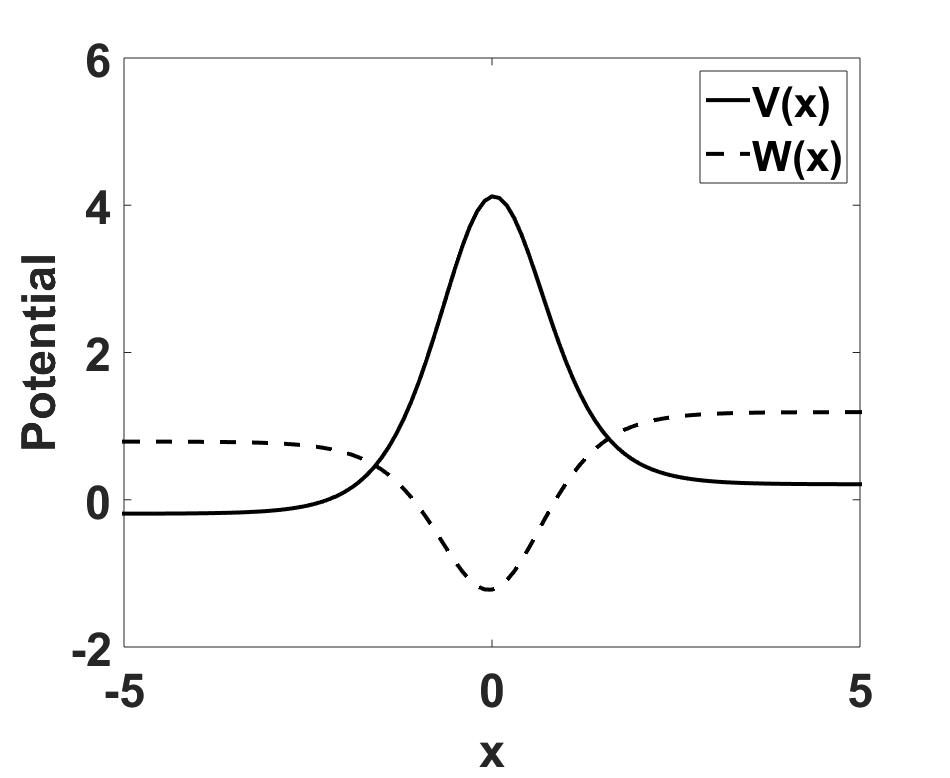}
\caption{}
\end{subfigure}
\begin{subfigure}[b]{0.32\textwidth}
\includegraphics[width=1\textwidth]{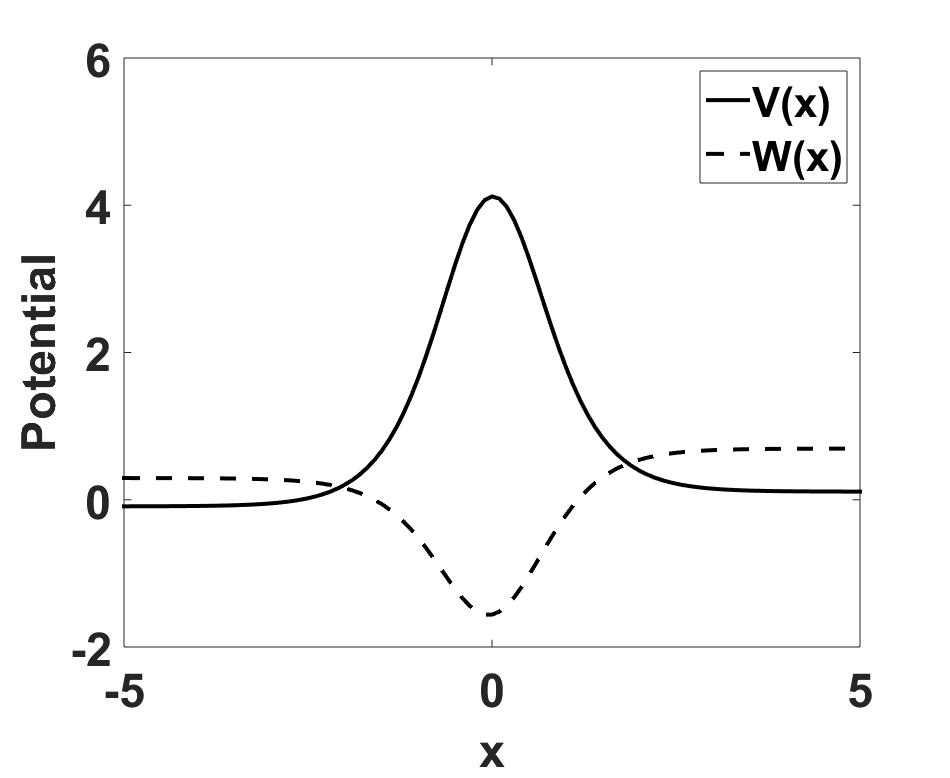}
\caption{}
\end{subfigure}
\begin{subfigure}[b]{0.32\textwidth}
\includegraphics[width=1\textwidth]{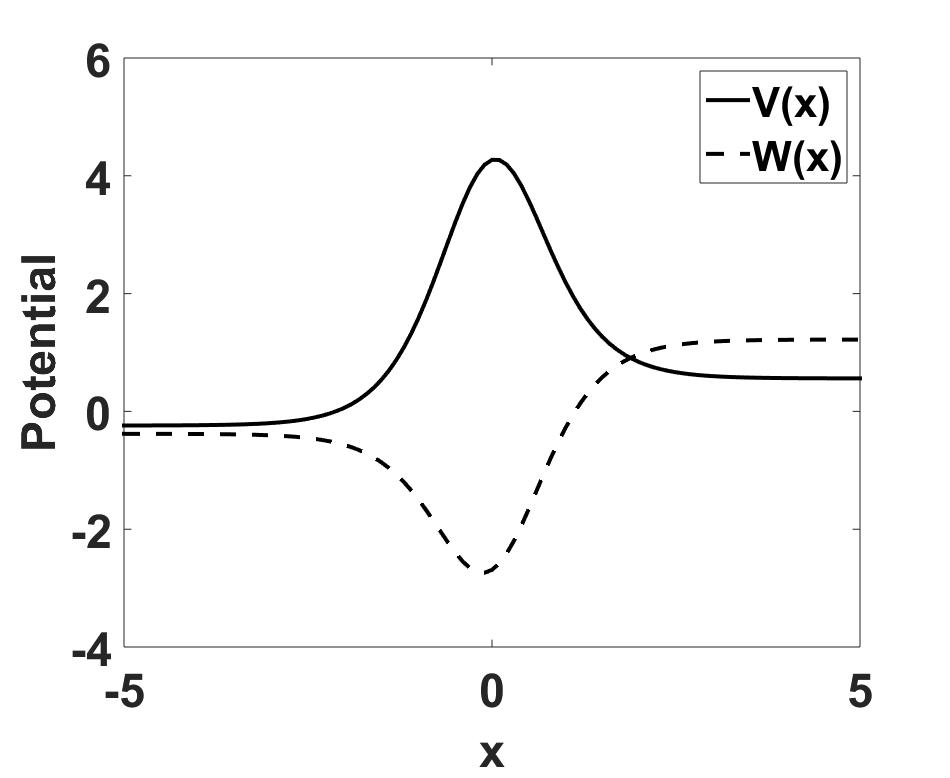}
\caption{}
\end{subfigure}
\begin{subfigure}[b]{0.32\textwidth}
\includegraphics[width=1\textwidth]{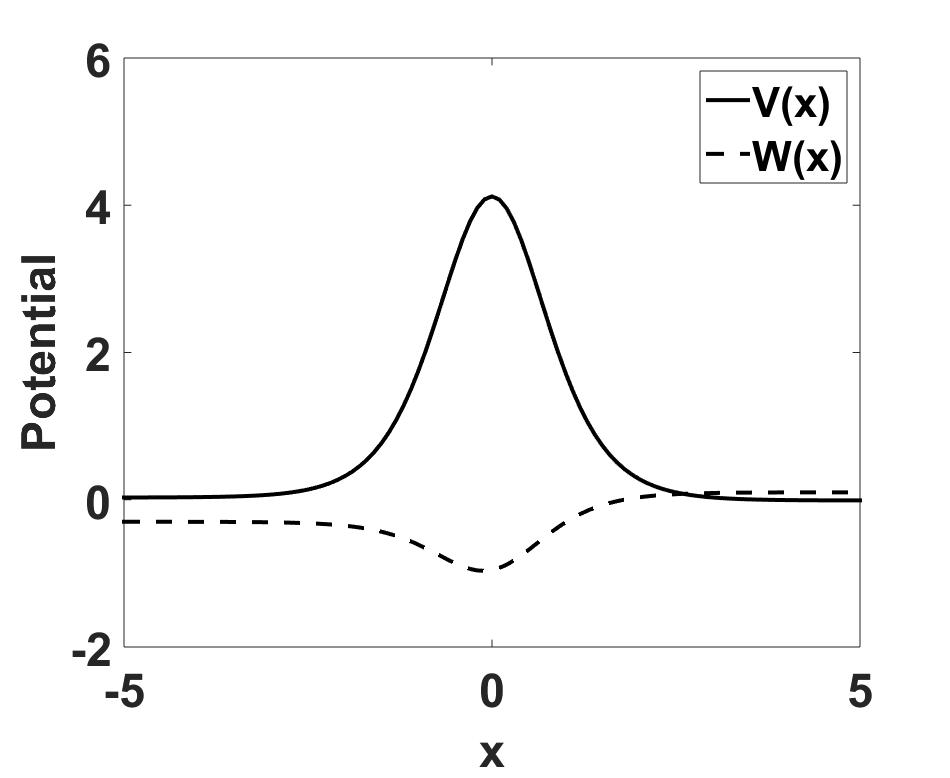}
\caption{}
\end{subfigure}
\begin{subfigure}[b]{0.32\textwidth}
\includegraphics[width=1\textwidth]{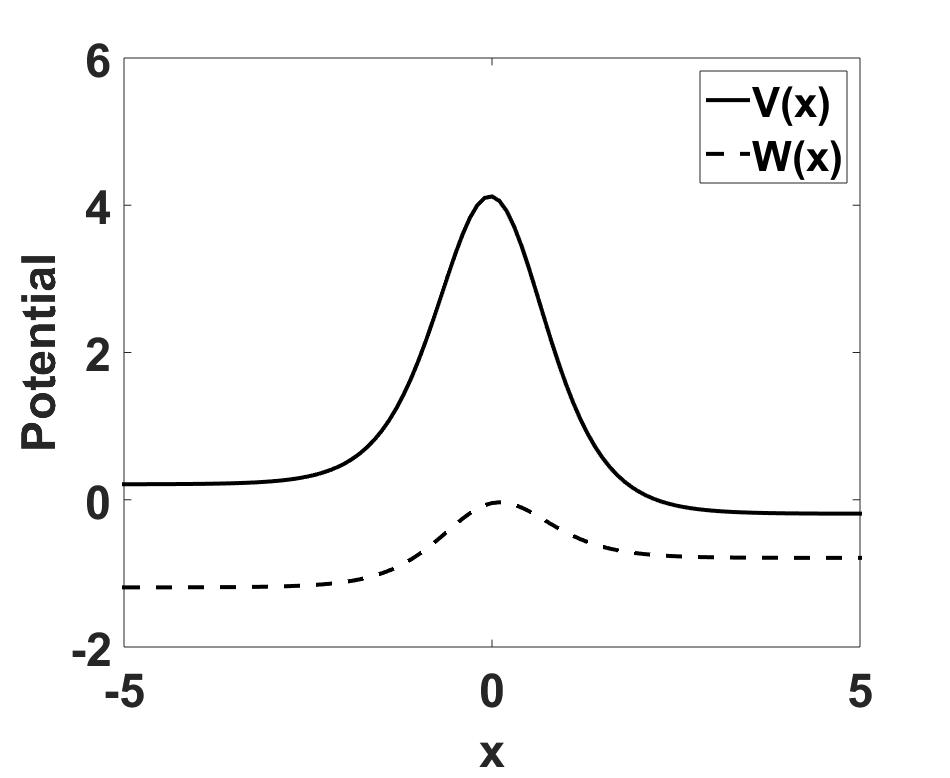}
\caption{}
\end{subfigure}
\caption{The nature of modified potential for self-defocusing mode with \textbf{(a)} $b=0.1$, $\alpha_2=-1 $ $\beta_{2}=0.1$;  \textbf{(b)} $b=0.1$, $\alpha_2=-1 $, $\beta_{2}=-0.1$;  \textbf{(c)} $b=0.1$, $\alpha_2=-0.5 $, $\beta_{2 }=-0.5$; \textbf{(d)} $b=0.4$, $\alpha_2=-0.5 $, $\beta_{2}=-1$; \textbf{(e)} $b=0.1$, $\alpha_2=0.1 $,  $\beta_{2}=-0.5$  and  \textbf{(f)} $b=0.1$, $\alpha_2=1 $,  $\beta_{2}=-0.5$.}
\label{fig8}
\end{figure}

The modified complex asymmetric potential for self-defocusing mode is given by 
\begin{subequations}\label{eq12}
\begin{align}
\label{eq12a} V(x)= & \left(a(a+1)+4 \alpha_1\right)\sech^{2}(x)-2\left(\frac{\alpha_2}{\alpha_1}\right)b\tanh(x)+\frac{b^2}{\alpha_1}, \\
\label{eq12b} W(x)=& 2b\tanh(x)+W_1\sech^2(x)+\alpha_2\left(\frac{b^2}{\alpha_1^2}-1\right),
\end{align}
\end{subequations}
where $W_{1}=2\alpha_{2}+(a(a+1)+2\alpha_{1})\left( \frac{\beta_{2}}{\beta_{1}}\right)$, $a$ and $b$ are positive real values for describing the strength of the potential. There is slight change in the potential for defocusing mode while compared with the earlier case.  Fig. \ref{fig8} shows the nature of potential (\ref{eq12}) for different parameter values.
Figures \ref{fig8}(a) and \ref{fig8}(b) indicate that, with the change in sign of $\beta_2$ near zero there is only a very small change in the shape of potential.  In Fig. \ref{fig8}(c), when $\alpha_2$ is increased from $-1$ to $-0.5$ the imaginary part of the potential shifts downwards. As observed in self-focusing mode, $b$ increases the asymmetry in the potential which is shown in Fig. \ref{fig8}(d).  Further, positive values for $\alpha_2$ shift the potential as in Figs. \ref{fig8}(e) and \ref{fig8}(f).  For $\alpha_2=1$ the imaginary part is inverted as depicted in Fig. \ref{fig8}(f) which was similar to self-focusing mode but real part is not inverted. The corresponding values of parameters are used for studying the stability and soliton evolution.

\subsection{Stationary soliton solution and linear stability analysis}
Extending our insight into self-defocusing nonlinearity, for which $\sigma=-1$, the exact soliton solution for Eq. (\ref{eq2}) is given by 	
\begin{eqnarray}\label{eq12d}
\phi(x)=\sqrt{-\frac{a(a+1)+2\alpha_{1}}{\beta_{1}}} \text{sech}(x) e^{\frac{ibx}{\alpha_{1}}},
\end{eqnarray}
where all the terms are already discussed in Sec. 3. Here also the condition for obtaining this solution is $\mu=\alpha_{1}$. The stationary solution in Eq. (\ref{eq12d}) is similar to self-focusing mode with an additional $i$ which leads to swapping of real and imaginary parts. The linear stability analysis for the above solution is studied as mentioned earlier and our results are plotted for different values of parameters. 

\begin{figure}[ht!]
\begin{subfigure}[b]{0.33\textwidth}
\includegraphics[width=1\textwidth]{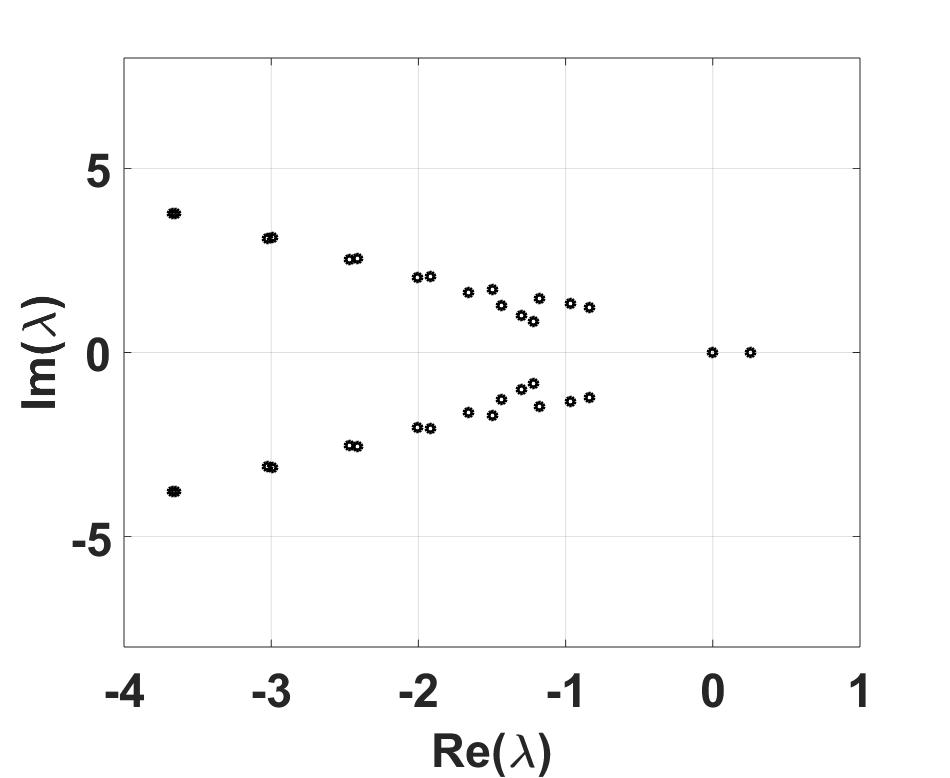}
\caption{}
\end{subfigure}
\begin{subfigure}[b]{0.33\textwidth}
\includegraphics[width=1\textwidth]{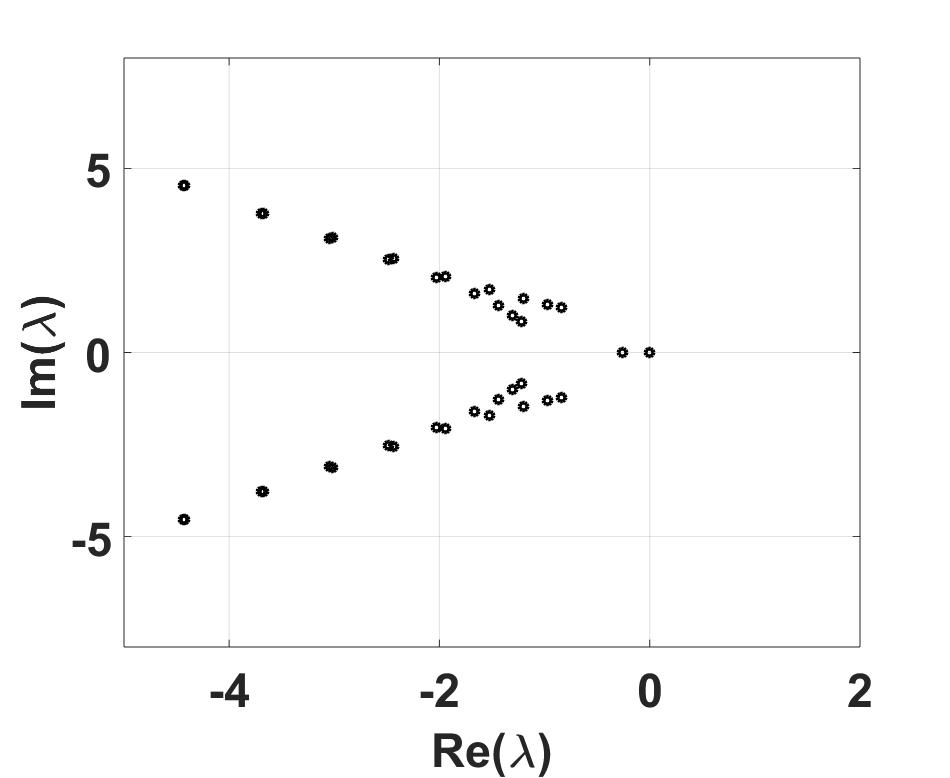}
\caption{}
\end{subfigure}
\begin{subfigure}[b]{0.32\textwidth}
\includegraphics[width=1\textwidth]{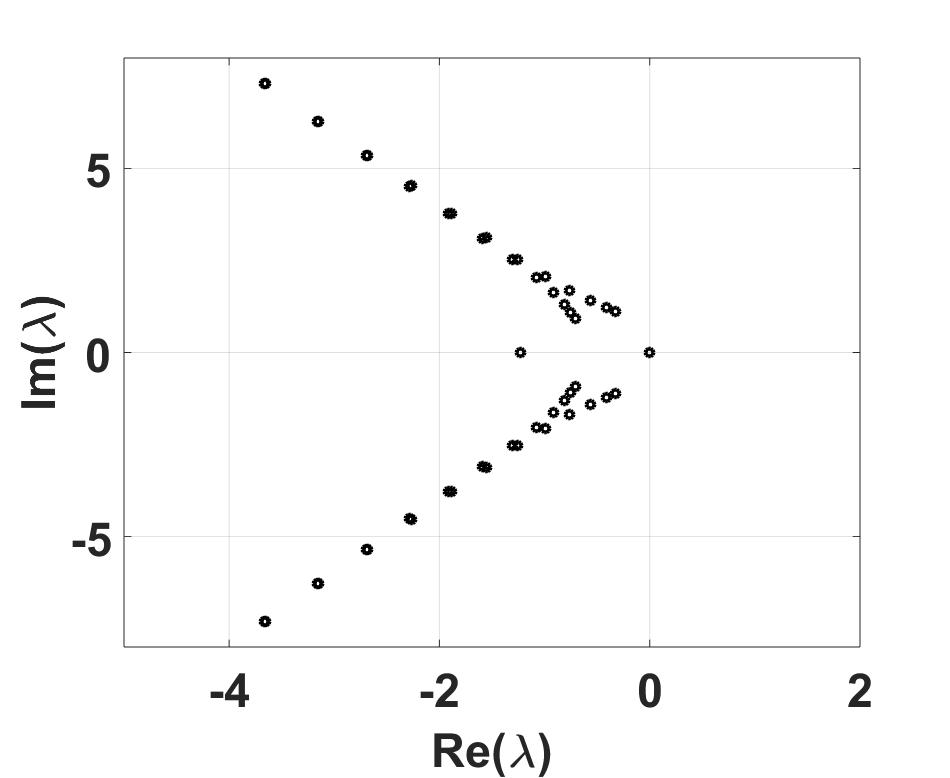}
\caption{}
\end{subfigure}
\begin{subfigure}[b]{0.33\textwidth}
\includegraphics[width=1\textwidth]{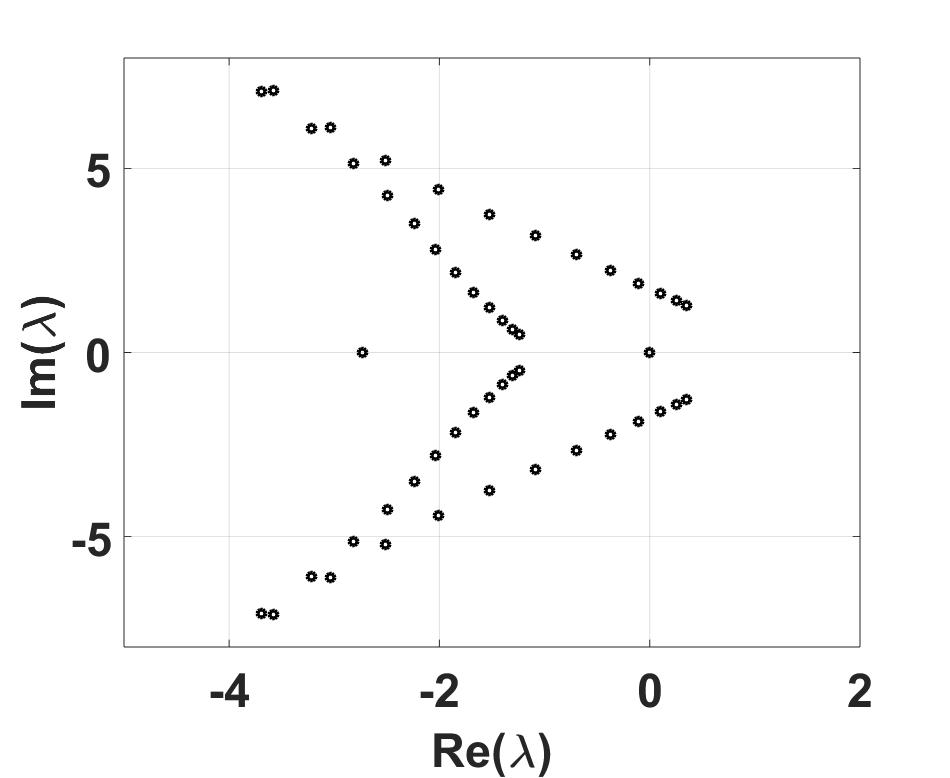}
\caption{}
\end{subfigure}
\begin{subfigure}[b]{0.33\textwidth}
\includegraphics[width=1\textwidth]{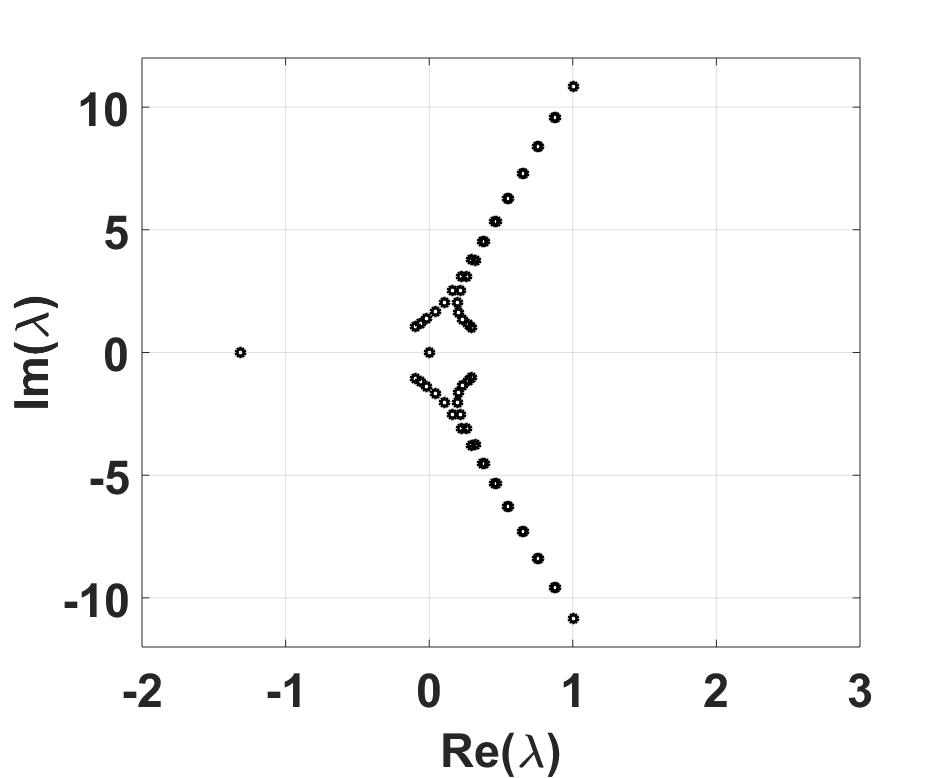}
\caption{}
\end{subfigure}
\begin{subfigure}[b]{0.32\textwidth}
\includegraphics[width=1\textwidth]{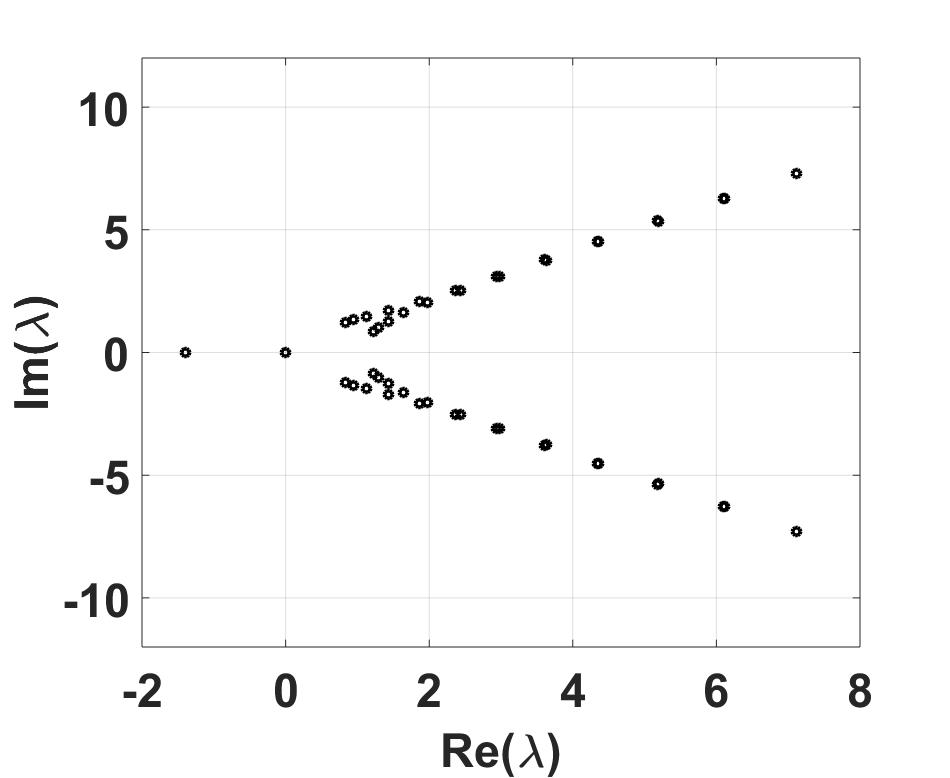}
\caption{}
\end{subfigure}
\caption{\textbf{(a)}$-$\textbf{(f)} Eigenvalue spectrum of stability analysis for the same order of parameters as in Fig. \ref{fig8}.}
\label{fig10}
\end{figure}

\begin{figure}[h!]
\begin{subfigure}[b]{0.32\textwidth}
\includegraphics[width=1\textwidth]{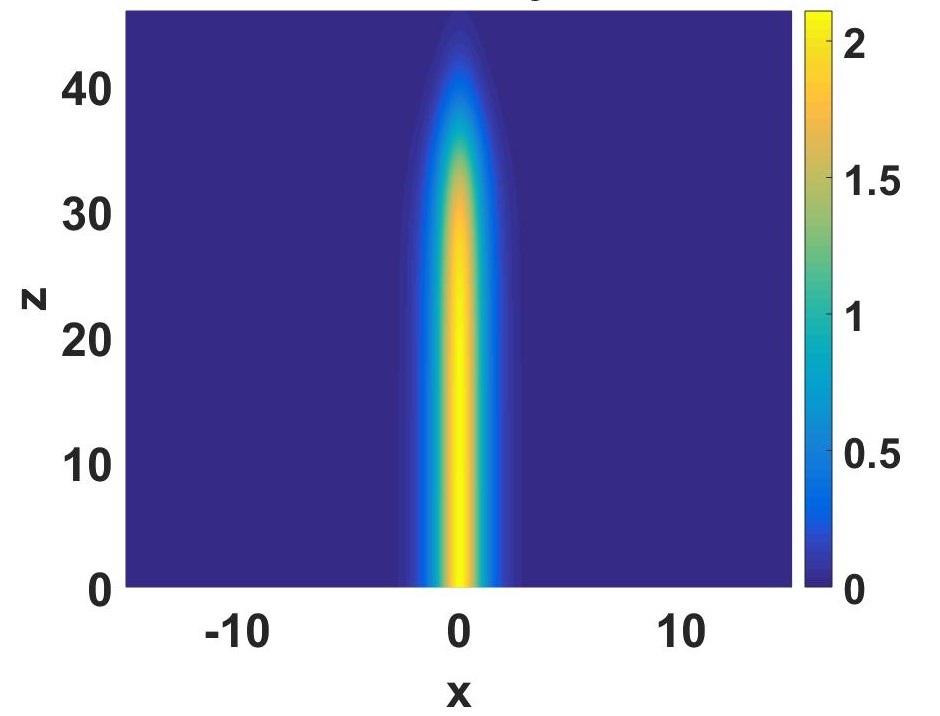}
\caption{}
\end{subfigure}
\begin{subfigure}[b]{0.33\textwidth}
\includegraphics[width=1\textwidth]{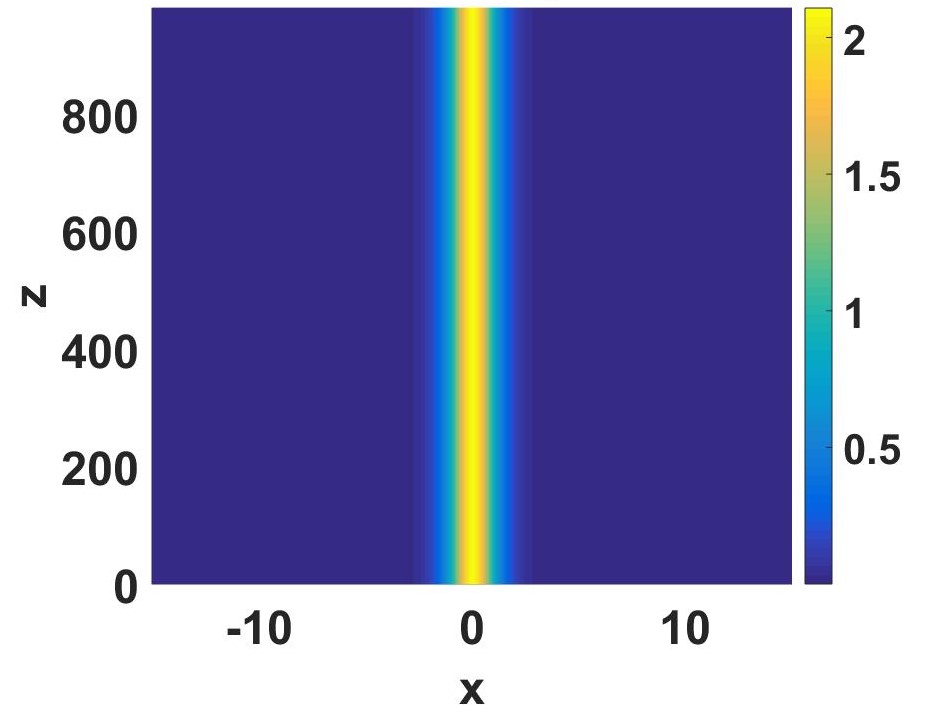}
\caption{}
\end{subfigure}
\begin{subfigure}[b]{0.32\textwidth}
\includegraphics[width=1\textwidth]{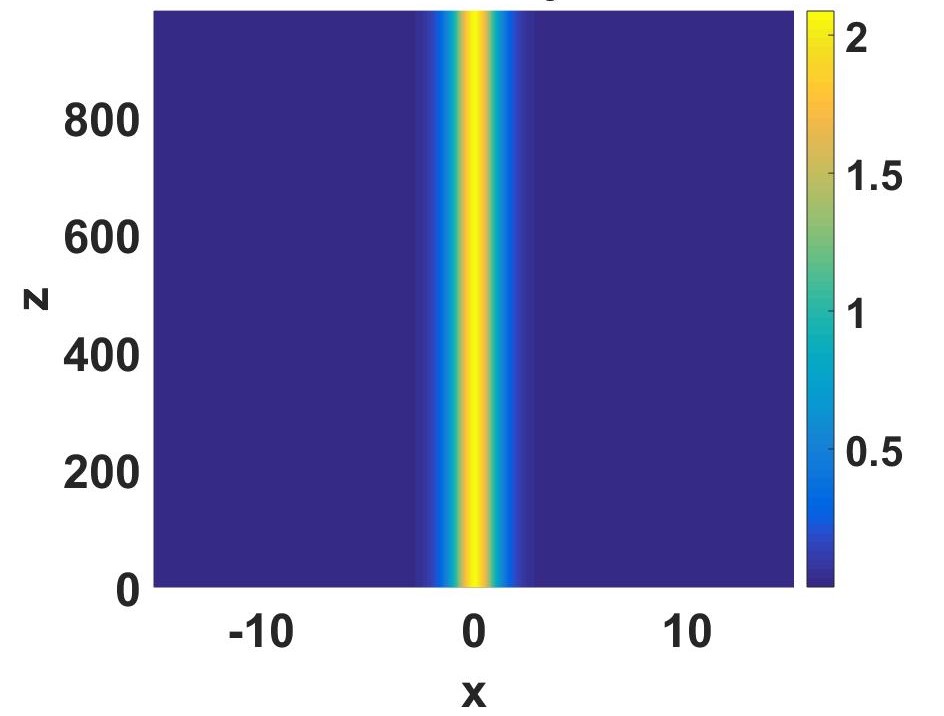}
\caption{}
\end{subfigure}
\begin{subfigure}[b]{0.33\textwidth}
\includegraphics[width=1\textwidth]{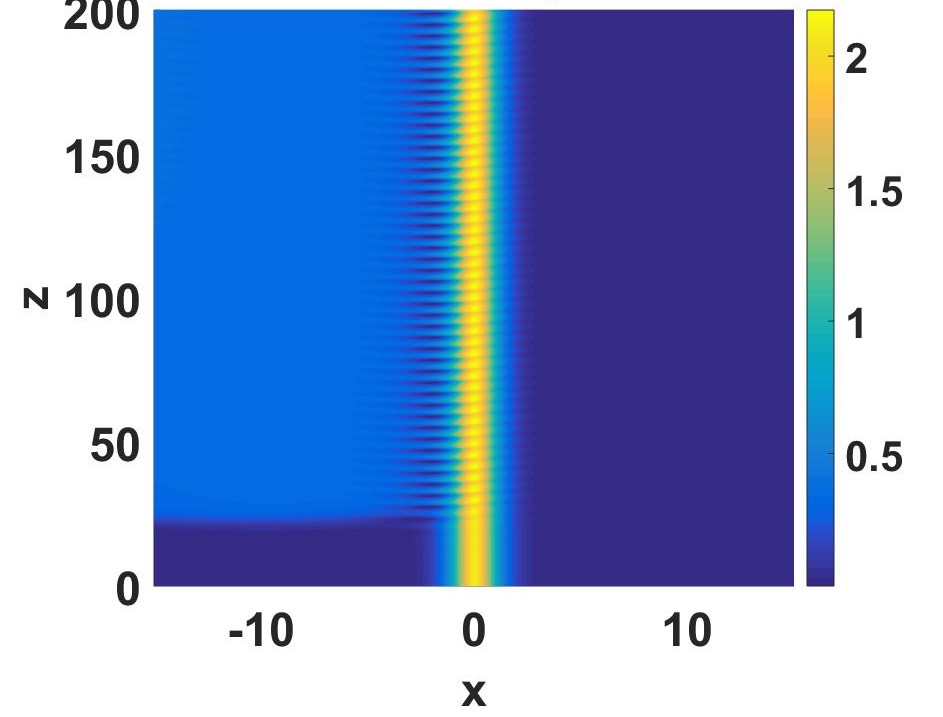}
\caption{}
\end{subfigure}
\begin{subfigure}[b]{0.33\textwidth}
\includegraphics[width=1\textwidth]{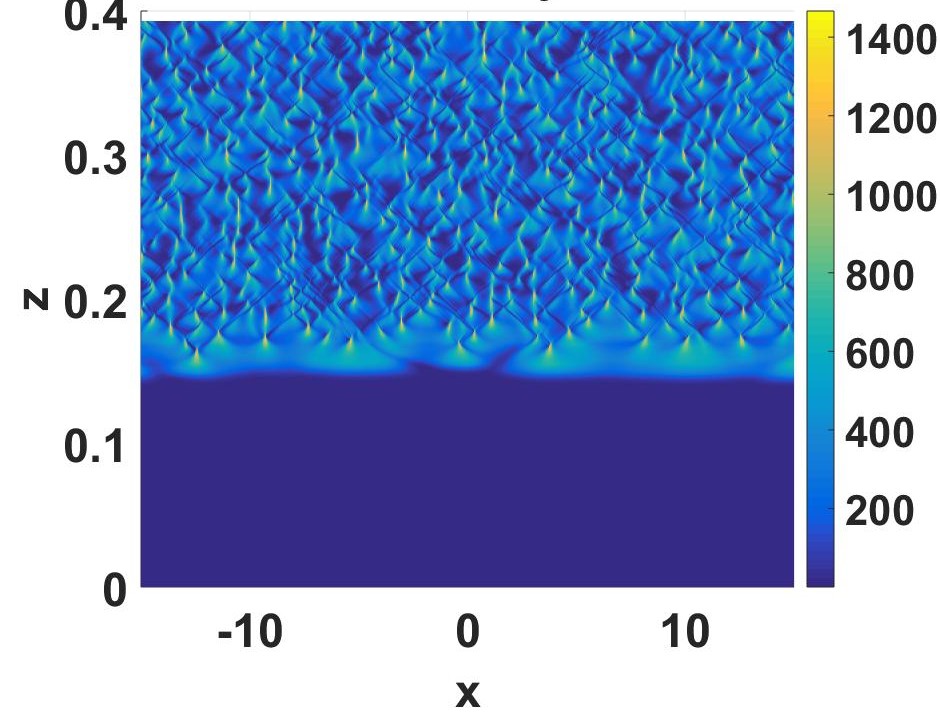}
\caption{}
\end{subfigure}
\begin{subfigure}[b]{0.32\textwidth}
\includegraphics[width=1\textwidth]{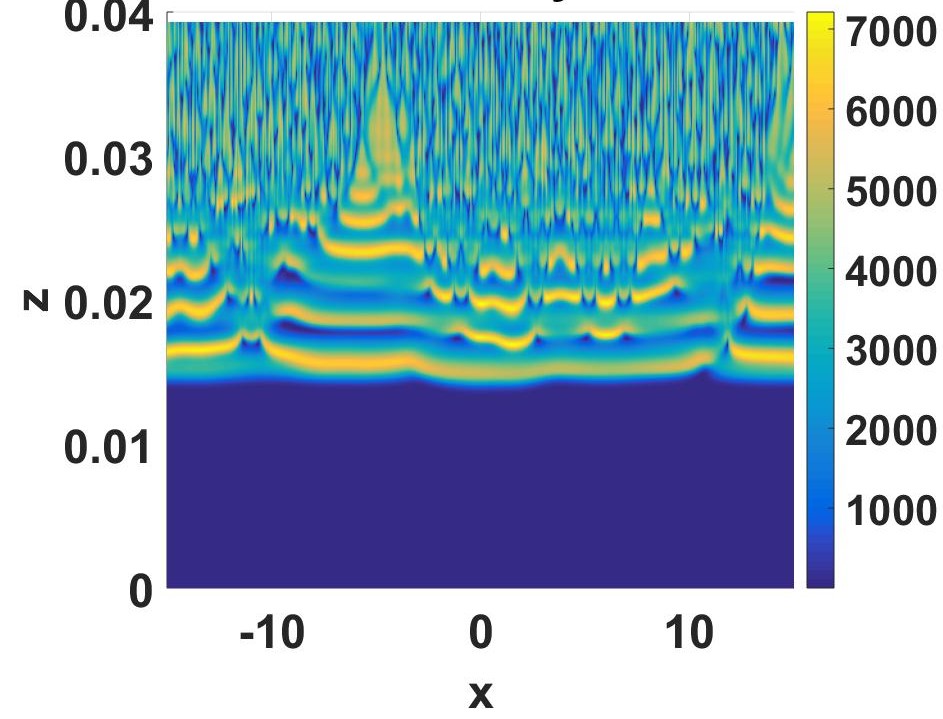}
\caption{}
\end{subfigure}
\caption{(Color online) \textbf{(a)}$-$\textbf{(f)} Intensity plot for stable and unstable evolution of soliton for the same order of parameters as in Fig. \ref{fig8}.}
\label{fig11}
\end{figure}

The stability analysis for self-defocusing nonlinear mode also shows the same sign pattern of $\alpha_{2}$ as earlier. Figure \ref{fig10}(a) has positive real part, as a result the soliton decays after a certain propagation distance. Figures \ref{fig10}(b) and \ref{fig10}(c) have negative real part and purely complex eigenvalues so that the soliton solutions are stable. The eigenvalues with positive real part such as Figs. \ref{fig10}(d)-\ref{fig10}(f) correspond to unstable solution. The evolution of stationary solution for the parameters obtained by stability analysis confirms the smoothness of soliton. Another important difference from self-focusing mode is that, if the value of $b$ is increased the system becomes unstable as shown in Fig. \ref{fig10}(d) where the real part of some eigenvalues are positive.

The evolution of soliton is given in Fig. \ref{fig11}. In case of self-defocusing nonlinear mode for positive values of $\beta_2$ the soliton travels for some time without any change in the amplitude and decays. This result is displayed in Fig. \ref{fig11}(a).   If $\alpha_2$ and $\beta_2$ values are negative with small value for $b$ then smooth propagation of solitons is possible as given in Figs. \ref{fig11}(b) and \ref{fig11}(c).  From Fig. \ref{fig11}(d), the soliton propagates with oscillation in amplitude and its instability is visible after $z=20$.  For positive value $\alpha_2$ the system becomes unstable as shown in Figs. \ref{fig11}(e) and \ref{fig11}(f), as in the case of self-defocusing mode.  

\subsection{Energy flow for exact soliton solution}
Similarly, for the self-defocusing case $(\sigma=-1)$, the gain or loss distribution of energy and the energy flux are given by 
\begin{subequations}
\begin{align}\label{eq14}
E=& -\frac{2b}{\alpha_{1} \beta_{1}}(a(a+1)+2\alpha_{1})\text{sech}^{2}(x)\text{tanh}(x),\\
j= & \frac{b(a(a+1)+2\alpha_1)}{\alpha_1\beta_1}\sech^2(x).
\end{align}
\end{subequations}
The energy flux and loss-gain distribution of the self-defocusing mode is same as the self-focusing mode. This implies that the stationary profile of dissipative soltions of both nonlinear modes are same. So the balance between the gain and loss is same. Here also the energy flow is independent of $\alpha_{2}$ and $\beta_{2}$ but dependent on the strength $(b)$ of the potential.

\section{Conclusion}
In our study, we modified the $\mathcal{PT}-$symmetric Rosen-Morse potential with NLS equation to a complex asymmetric potential with CGL equation.  Exact dissipative soliton solutions for this potential along with stability analysis and evolution of soliton solution for different parameter ranges are analyzed for both self-focusing and self-defocusing modes. As expected the dissipative solitons are stable for only certain range of parameters and do not possess in continuous families which are parameterized by the propagation constant. Stable soliton solutions are apparently possible for $\alpha_{2}<0$.  We have also investigated the energy flow in the potential and observed that the direction of energy flow in self-focusing and self-defocusing nonlinear modes are same. Analyzing the possibilities of stable soliton solutions can be further developed into experimental realizations in future.

\section*{Acknowledgments}
The authors would like to thank the Referees for their valuable comments and suggestions.  KM wishes to thank the Department of Science and Technology - Science and Engineering Research Board, Government of India, for providing National Post Doctoral fellowship under the Grant No. PDF/2016/001620/PMS.

\end{document}